\documentclass[aps,prd,12pt,showkeys,superscriptaddress,nobibnotes,longbibliography]{revtex4-1}

\usepackage{mathrsfs,graphicx,rotating,amsmath,amsfonts,mathtools,booktabs,amssymb,wasysym}
\usepackage{hyperref}
\usepackage[table,xcdraw,dvipsnames]{xcolor}
\usepackage{braket}
\usepackage{MnSymbol}

\usepackage{subfig}
\usepackage{caption}
\hypersetup{
     colorlinks   = true,
     citecolor    = blue,
     urlcolor     = blue,
     linkcolor    = blue
}

\newcommand{\med}[1]{\langle #1\rangle}

\newcommand{\LDC}{\Lambda_{\rm D}}
\newcommand{\SU}{\,{\rm SU}}
\newcommand{\Tr}{\,{\rm Tr}}
\newcommand{\adc}{\alpha_{\rm D}}
\newcommand{\MDM}{M_{\text{DM}}}

\newcommand{\be}{\begin{equation}}
\newcommand{\ee}{\end{equation}}

\newcommand{\PRL}{Phys. Rev. Lett.}

\newcommand{\dd}{\,\text{d}}

\makeatletter

\def\hhref#1{\href{http://arxiv.org/abs/#1}{arXiv:#1}}
\usepackage{xstring} 
\newcommand{\hhrefq}[1]{\IfSubStr{#1}{:}{\href{http://inspirehep.net/search?ln=en&ln=en&p=#1&of=hb&action_search=Search&sf=&so=d&rm=&rg=25&sc=0}{InSpires:#1}}{\hhref{#1}}}

\def\art{\@ifnextchar[{\eart}{\oart}}
\def\eart[#1]#2#3#4#5#6{{\rm #2}, {\em #3 \bf #4} {\rm (#6) #5} ({\em #1})}
\def\article{\@ifnextchar[{\earticle}{\oarticle}}
\def\oarticle#1#2#3#4#5#6{{\rm #1}, {``#6''}, {\rm #2 #3 (#5) #4}}
\def\earticle[#1]#2#3#4#5#6#7{{\rm #2}, {``#7''}, {\rm #3 #4 (#6) #5}  [\hhrefq{#1}]}
\def\hepart[#1]#2{{\rm #2, \sl#1}}
\def\heparticle[#1]#2#3{#2, { ``#3''} [\hhrefq{#1}]}

\colorlet{ins}{blue} 
\colorlet{del}{red}

\usepackage{xcolor} \usepackage{ulem} \usepackage{changebar}

\begin{document}
{\hfill CP3-Origins-2019-21 DNRF90\hfill}

\title{The Thermal History of Composite Dark Matter}

\author{Nicola Andrea Dondi}
\thanks{{\scriptsize Email}: \href{mailto:dondi@cp3.sdu.dk}{dondi@cp3.sdu.dk}; {\scriptsize ORCID}: \href{https://orcid.org/0000-0002-6971-2028
}{0000-0002-6971-2028}}
\affiliation{$\text{CP}^3$-Origins, University of Southern Denmark, Campusvej 55, 5230 Odense, Denmark}

\author{Francesco Sannino}
\thanks{{\scriptsize Email}: \href{mailto:sannino@cp3.sdu.dk}{sannino@cp3.sdu.dk}; {\scriptsize ORCID}: \href{http://orcid.org/0000-0003-2361-5326}{0000-0003-2361-5326}}
\affiliation{$\text{CP}^3$-Origins, University of Southern Denmark, Campusvej 55, 5230 Odense, Denmark}    

\author{Juri Smirnov}
\thanks{{\scriptsize Email}: \href{mailto:smirnov@cp3.sdu.dk}{smirnov@cp3.sdu.dk}; {\scriptsize ORCID}: \href{http://orcid.org/0000-0002-3082-0929}{ 0000-0002-3082-0929}}
\affiliation{$\text{CP}^3$-Origins, University of Southern Denmark, Campusvej 55, 5230 Odense, Denmark}

\begin{abstract}
\noindent
\begin{center}
\textbf{ABSTRACT}
\end{center}
\noindent  We study the thermodynamic history of composite Dark Matter models. We start with classifying the models by means of the symmetries partially protecting the composite Dark Matter decays and constrain their lifetimes. For each model, we determine the impact of number-changing and number-conserving operators on its thermal history. We also develop the analytic formalism to calculate the asymptotic abundance of stable relics. We show how the relative strength between number- changing and number-conserving interactions together with the dark plasma lifetime affect the thermal fate of the various composite models. Additionally, we show that the final dark relic density of composite particles can be diluted due to an entropy increase stemming from dark plasma decay. Finally, we confront the models with experimental bounds. We find that indirect detection experiments are most promising in testing this large class of models.
\end{abstract}
 \maketitle

\newpage

\tableofcontents

\newpage


\section{Introduction}

The nature of dark matter is one of the most fascinating questions that remain to be addressed in particle physics. Several ideas have been put forward that can be broadly classified according to whether dark matter emerges as an elementary particle or a composite one made by more fundamental matter. Both possibilities have been explored in the literature \cite{Ge:2019voa, Berges:2019dgr, Blennow:2019fhy, Arcadi:2019lka,Cirelli:2018iax,1503.03066,1105.5431,1308.4130,1511.04370,1904.12013, Roszkowski:2017nbc, 1411.3739, DiracDM, DirectVSIndirect, LeptophillicDM, Khlopov:2011tn, Khlopov:2014bia, 1607.07865}. 

Among the proposed models, one class is particular in the following sense. In a model in which the dark matter particle is in thermal equilibrium in the early universe, the asymptotic relic abundance is uniquely determined by its decoupling process from the thermal plasma. The best studied scenario is based on the process $\text{DM} + \text{DM} \rightarrow \text{SM} + \text{SM}$, the WIMP annihilaion~\cite{Steigman:1984ac, SteigmanDasguptaBeacom}. This process provides us also with a target annihilation cross section, which can be searched for in indirect detection experiments. 
However, other processes for dark matter decoupling have been considered as well. Number changing interactions among dark matter particles only $\text{DM} + \text{DM}  + \text{DM} \rightarrow  \text{DM}  + \text{DM}$ lead to the SIMP freezeout~\cite{1402.5143}, and a mixed topology $\text{DM} + \text{DM}  +  \text{SM} \rightarrow  \text{DM}  + \text{SM}$ to the Co-SIMP freezeout~\cite{2002.04038}. The advantage of the thermal models is their predictivity, which gives us guidelines for testing those hypothesis. 

In this paper, we focus on the thermal history and ultimate fate of some of the prime candidates of composite dark matter. A generic property of composite models of dark matter is that, at low energies, they feature a complex multicomponent structure~\cite{1312.3325}. In particular, in the confined phase the lightest composite particles can kinetically decouple from the Standard Model plasma and evolve as an independent dark thermal bath. This can significantly affect the relic abundance and late time target cross section and we will elucidate this issue in several scenarios arising from a dark composite model. Our analysis is based on features of the low energy theory common to many confining models, attaining a certain level of generality. In some explicit examples, we will, however, employ a minimal setting consisting of $SU(N)$ gauge theories with fundamental fermions.

Composite dark matter arising in confining gauge theories is a well-motivated particle physics scenario to explain the observed missing mass problem. Its particular advantage lies in the fact, that it provides an explanation to the long dark matter lifetime.  The explanation is analogous to the Standard Model proton stability, which can be understood in terms of an accidental global symmetry, the baryon number. In the confining gauge theories, new accidental global symmetries arise and lead to long-lived relic particles~\cite{AccidentalDM,ColoredDM,1811.06975}. 

The paper is structured as follows. In section~\ref{sec:exsummary} we give a brief summary of our philosophy and main findings. In section~\ref{sec:candidates} we identify the composite dark matter candidates arising in gauge theories. We classify them according to the symmetries protecting their decays, as suggested in~\cite{1312.3325, AccidentalDM,1811.06975, 1811.03608}. We further discuss higher-dimensional operators inducing composite dark matter decay that can emerge in a full ultraviolet theory including gravity, as suggested in Ref.~\cite{AccidentalDM}.  We use these operators to estimate the composite dark matter candidates' lifetime.  

Section~\ref{sec:thermo}  is devoted to the thermal history of the composite dark sector. For each model, we elucidate the impact of both number-changing and number-conserving operators~\cite{Canibal1,Canibal2}. In section~\ref{sec:freeze-out} we construct solutions to the Boltzmann equations allowing us to determine the asymptotic abundance of stable relics. 

This preparatory phase is then exploited in section~\ref{sec:models} to determine the relic abundance in each model. Here we provide analytic expressions for the asymptotic relic abundance stemming from each scenario. We observe that the relative strength between number-changing and conserving interactions determines the thermal fate of the various composite models and further depends on the dark plasma lifetime. 

We find that the final composite dark relic density can be diluted due to an increase in entropy stemming from dark plasma decay, as pointed out in~\cite{BaryonDM, 1811.06975}.  We constrain the models with strongly coupled bound states~\cite{AccidentalDM} and weakly coupled bound states~\cite{BaryonDM} via the experimentally observed relic density. The entropy ratio between the standard model and the dark sector is another parameter affecting the ultimate relic abundance which we fix by assuming the two plasma to be in thermal equilibrium in the early universe.  

Section~\ref{sec:constraints} confronts the model predictions with experiments such as H.E.S.S. and Fermi-LAT. The details of the experimental signatures depend on the concrete Standard Model quantum numbers chosen for the dark quarks, a classification of viable quantum number assignments can be found in Ref.~\cite{AccidentalDM}. Finally, we offer our conclusions in section~\ref{sec:conclusions}. 

In appendix~\ref{appendixA} and~\ref{appendixB}  we offer a glossary of multi-fluid thermodynamics and in appendix~\ref{appendixC} we summarize the Boltzmann equation for number-changing processes. 

 
 \section{Executive Summary for the Busy Reader}
\label{sec:exsummary}

As we will show in this work, confining dark sectors provide a very rich and interesting phenomenology and phase structure, despite being based on models with only a few ingredients and parameters. Those parameter are, in the UV regime, the number of gluons, based on the chosen confining gauge group, the number, mass and quantum number assignments of dark quarks and the strength of the dark gauge coupling. This coupling strength can be related to the confinement temperature scale, the relation of which to the quark masses determines the dynamics of the system. 

One central question of any dark matter model building endeavor is dark matter stability. This question, can be naturally addressed in a confining dark sector, since it has particles which are protected by accidental global symmetries, just like the proton in the Standard Model. Generically, as we will discuss, the considered theories have light particles with lifetimes shorter than the age of the universe and long-lived heavier states. For this reason, thermal freezeout in such a dark sector shows interesting complex phenomena.  In particular, the bath of the lighter dark particles, which we also call the dark plasma, can undergo self-heating effects and significantly change the freezeout of the heavier long-lived dark matter candidates. Note that from now on, we denote by freezeout the deviation of interaction rates from the thermal interaction rate, leading to particle number conservation by the considered process. 

As a template model, we will focus on the $SU(N_{DC})$ theory with vector-like fermions in the fundamental representation. We additionally assume that there is at least one dark quark in the spectrum, which carries both dark sector and standard model quantum numbers. This setup provides a thermal-contact bridge between the sectors at large temperatures. It thus unavoidably sets an upper bound on the entropy ratio between the Standard Model sector and the dark sector $\zeta = s_{\rm SM}/s_D < g_{\rm SM}\approx 106$.  

We demonstrate the freezeout in different mass hierarchies of the model. It is particularly interesting that larger target cross sections, compared to the standard scenario with $\langle \sigma v_{\rm rel. }\rangle \sim 10^{-26} \text{cm}^3/s$, are expected. Furthermore, decreasing the $\zeta$ parameter introduced above by postulating dark sectors with more degrees of freedom, increases the expected late time annihilation cross sections, making the scenarios easier accessible. This allows to systematically test the dark sectors with thermal dark matter in indirect detection experiments. 

Very broadly the systems can be classified in the following way:
\begin{itemize}
\item Strongly coupled dark baryons as dark matter candidates, with light dark pions comprising the dark plasma. 
\item Weakly coupled dark baryons as dark matter candidates, with light dark glueballs being the dark plasma particles.
\end{itemize}

Lastly, we also discuss the possibility of long-lived dark plasma particles. It appears that to be consistent with existing experimental limits, they should comprise a subdominant dark matter fraction. This naturally leads to multi-component dark matter sectors with possible implications for structure formation in the dark sector.

 
 \section{Composite Dark Matter Candidates}
\label{sec:candidates}
A particle description of dark matter (DM)  requires a lifetime above $\tau > 10^{26} - 10^{28} \text{ s}$ \cite{DM decay, 1612.05638} which is orders of magnitude times the age of the universe ($10^{17}\text{s}$). A near-exact symmetry is a time-honored way to ensure long lifetimes.  A famous example is the baryon symmetry that protects the proton from decaying.  For this reason, we investigate here models of composite DM  stemming from the dynamics of Yang-Mills gauge theories featuring fermionic matter. In these models, we identify potentially long-lived composite relics and systematically investigate their cosmological consequences.

 Historically, models of composite DM, such as the ones inspired by Technicolor~\cite{Nussinov:1985xr, hep-ph/0603014, hep-ph/0608055, 1105.5431}, linked the baryon asymmetry to a potential DM asymmetry in order to achieve the experimentally observed dark relic density.  Another class of composite DM models was introduced to generate the desired relic density via number-changing operators in the dark-pion sector~\cite{Hochberg:2014kqa}. Here we consider composite DM models in which the primary source for the observed relic density is due to a thermal production mechanism.


\subsection{The Composite Framework} \label{sec:model}

As illustrative example we consider a $SU(N_{DC})$ gauge theory with $N_f$ fundamental Dirac fermions $\chi$ and $N_f$ anti-fundamental Weyl fermions $\psi$, $N_{DC}$ being the number of dark colors. The Lagrangian is:
\begin{equation}
\mathcal{L} = -\frac{1}{4} G^{\mu\nu}G_{\mu\nu} + \sum_{i=1}^{N_f} \bar{Q}_i( i \gamma^{\mu} D_{\mu} - M_Q )Q_i   ~.
\end{equation}
The well known local and global symmetries of the dark quarks are summarised in Tab. \ref{tab:DMQN}.
For sufficiently small $N_f$ this model confines at a scale $\Lambda_D$ and the chiral symmetry is spontaneously broken. The breaking pattern $SU(N_f)_L \otimes SU(N_f)_R \rightarrow SU(N_f)_V $ has an order parameter
\begin{equation}
\langle \bar{Q}_i Q_j \rangle = \Lambda_D^3 \delta_{ij} ~.
\end{equation}
If we assume an hierarchy of scales as
\begin{equation}
\Lambda_D = 4\pi f_{\pi} \gg M_Q \quad \forall i ~,
\end{equation} 
the effective low energy theory is described in term of a unitary matrix $U(x) = \exp{ i \Pi(x)/f_{\pi}}$ where $\Pi = \sum_a T^a \pi^a$ are Goldstone bosons associated with the breaking patterns.  Their dynamics at lowest order in derivatives is described by:
\begin{equation}
\mathcal{L} = \frac{f_\pi^2}{8} \Tr[ \partial_{\mu}U\partial^{\mu}U^{\dagger}] + \frac{f_\pi^3 M_Q}{4} \Tr[ U^{\dagger} + U ] ~,
\label{eq:ChiralLagrangian}
\end{equation}
where $M$ is a matrix in flavour space that parametrises the explicit breaking of chiral symmetry due to quark masses. The lowest order operator providing number-changing processes is a $5$-point interaction, the so-called Wess-Zumino-Witten term, present for specific breaking patterns \cite{Hochberg:2014kqa}. In our case this reads:
\begin{equation}
\mathcal{L}_{WZW} = \frac{N_{DC}}{240\pi^2 f_\pi^5} \epsilon^{\mu\nu\rho\sigma} \text{Tr}\left[ \Pi \partial_{\mu} \Pi \partial_{\nu} \Pi \partial_{\rho} \Pi \partial_{\sigma} \Pi \right] ~.
\end{equation}

Because much is known for $N_{DC}=3$ either via experiments or via lattice simulations we use it as our benchmark model. 


\begin{table}
 $$ \begin{array}{cc|ccc} \nonumber
 & [\SU(N_{DC})]  & \SU (N_f)_L & \SU(N_f)_R & {\rm U}(1)_{V}\\ \hline
Q      & \Box  & \Box & 1 & 1\\ \hline
\bar{Q}      & \bar{\Box }  & 1 & \Box & -1\\ \hline
\end{array}
$$
\caption{\label{tab:DMQN}  Field content of the lightest fermions in the model. }
\end{table}

On general grounds, part of the dark building blocks will carry standard model (SM) quantum numbers that can bring into thermal equilibrium the dark and the SM sectors.  

Depending on whether  $M_Q \ll \LDC$ (strongly coupled regime) or $M_Q \gg \LDC$ (weakly coupled regime) the composite states will behave differently. In the latter case the low energy theory is a pure Yang-Mills with the confinement scale that can be estimated to be: 
\begin{align}
\frac{\LDC}{ M_Q}  \approx \exp{ \left( - \frac{6 \pi}{11 N_{DC} \, \adc(M_Q) } \right)}  \, ,
\end{align}
with $ \adc(M_Q) = g^2_D/4\pi$ the dark coupling strength. 
 
 Because of either the presence of dark fermion mass or the generation of the Yang-Mills confining scale none of the lightest composite particles (LCP) is massless. Respectively for $M_Q \ll \LDC$ and $M_Q \gg \LDC$ the LCP are:
 
\begin{itemize}
\item The near  Nambu-Goldstone bosons of the broken global symmetry (pions), with dynamics described above. 
\item The dark gluon bound states (glueballs), with the lightest state being a scalar singlet.  
\end{itemize} 

Another class of interesting composite states is constituted by composite baryons. The reason being that, even though they are not LCPs,  the dark baryon number is protected. 
As a teaser, we notice that in both limits the LCPs will feature number-changing operators generating intriguing thermodynamics. 

\subsection{Composite Particles and their Lifetimes}

Here we summarise the composite particle spectra and discuss their lifetimes due to the possible breaking of the accidental global symmetries via higher-order operators~\cite{AccidentalDM}. The spectrum and operators are summarised for the reader's convenience in  Table~\ref{tab:lifetimes}. 

\begin{table}
 $$ \begin{array}{c|cccc} \nonumber
 \hbox{State} & \hbox{Symmetry}  & \hbox{Decay Operator} & \hbox{Decay Operator}  & \hbox{allowed mass}\\ \hline
   & \text{(if applicable) } & \text{(fundamental fields) } & \text{(on EFT level) }   & M_\text{max} \\ \hline
    Q^{N_{DC}}  &  \hbox{Dark Baryon Number}& \frac{Q^{N_{DC}}\,f}{\Lambda_{\mathcal{B} \!\!\! /}^{3/2 (N_{DC}-5/3)} } &   \Lambda_{\mathcal{B} \!\!\! /} \,\,   \left(\frac{\LDC}{\Lambda_{\mathcal{B} \!\!\! /}}\right)^{3/2(N_{DC}-1)} \mathcal{B} f & \gtrapprox 300 \text{ TeV} \\ \hline
    \bar{Q} Q &  \hbox{G-parity} &  \frac{ \alpha_\text{EM} \, c_{i j} \, \bar{Q}_i\,F_{\mu \nu} \sigma^{\mu \nu}\,Q_j}{\Lambda_{G \!\!\!\! /}} & \frac{\LDC^2 }{\Lambda_{G \!\!\!\! /}}  \, \alpha_\text{EM} \, c_{i j} M_{i j} \,F_{\mu \nu} \sigma^{\mu \nu} & \approx 10 \text{ MeV} \\ \hline
 \bar{Q_i} Q_j & \hbox{Species Number} & \frac{ c_{i j} \, \bar{Q}_i\,Q_j H^\dagger H}{\Lambda_{S \!\!\!\! /}}  & \frac{\LDC^2 }{\Lambda_{S \!\!\!\! /}}  c_{i j} M_{i j} \, H^\dagger H     & \approx  5 \text{ MeV} \\ \hline
G_{\mu \nu}G^{\mu \nu} & - & \frac{ G_{\mu \nu}G^{\mu \nu} H^\dagger H}{\Lambda^2} & \frac{ \LDC^3}{\Lambda^2}\,S\, H^\dagger H  & \approx  50 \text{ TeV} \\ \hline
\end{array}
$$
\caption{\label{tab:lifetimes} Composite particles and their maximal masses, compatible with DM life-time limits. We list the decay operators which lead to the decay of the DM candidates to SM fields with the lowest possible dimension, as they will be most efficient. $F^{\mu\nu}$ is the field strength tensor of $U(1)$ electromagnetic, while $f$ are SM matter fields. It is assumed that the symmetry violation scale associated with the decay operators is at most  $\Lambda \approx M_\text{Pl}$, since the global symmetries are expected to break down at least at the quantum gravity scale. Here $\Lambda_{\mathcal{B} \!\!\! /}, \Lambda_{G \!\!\!\! /}$ and $ \Lambda_{S \!\!\!\! /}$ are the scales at which baryon number, G-parity and species number are violated, respectively.}
\end{table}

As mentioned above, among the composite particles we have dark baryons $  Q^{N_{DC}}  \sim \Lambda_D^{\frac{3}{2} (N_{DC} -1)} \mathcal{B} $ (for this case we consider $N_{DC}$ odd), dark mesons $   \bar{Q} Q \sim \Lambda_D^{2}\, \pi$ and last but not the least dark glueballs $G_{\mu \nu}G^{\mu \nu}  \sim \Lambda_D^{3} S$.  Their decays can occur via higher dimensional operators such as the ones envisioned in Table~\ref{tab:lifetimes} and generated by new dynamics or gravitational interactions. We also introduced the interpolating low energy fields with the help of dimensional analysis.

Due to the dark flavor structure of the models, the mesons can carry further accidental symmetries as well.  Examples are: The species number, which would make  $\pi_{\pm}$ stable in the SM were not for the presence of weak interactions that violate this symmetry; Generalised G-parity~\cite{1005.0008}, which would stabilize $\pi_0$ but it is broken by the chiral anomaly that induces its decay into photons. Additionally, new physics can lead to induced higher-dimensional operators that further break these symmetries. These, however, are suppressed by the scale of the new physics.  
Using dimensional analysis for the dark sector and constraining ourselves to dimension five operators we  naively estimate the decay rates to scale as
  $\Gamma \propto m_\pi^3/\Lambda_{\mathrm NP}^2$ for the dark mesons in terms of the new physics scale $\Lambda_{\mathrm NP}$. Intriguingly even for $\Lambda_{\mathrm NP} \approx M_\text{Pl}$ one discovers that the experimentally required lower bound on the dark lifetime induces an upper bound  \cite{DM decay, 1612.05638} of $m_\pi \lessapprox 10 \text{ MeV}$ for this type of DM.

In general, within this framework, several relics can be long-lived on cosmic scales. Their abundances have to satisfy the overclosure bound $\sum_i \Omega_i \leq \Omega_\text{DM}$. Particles with lifetimes shorter than the current life-time bound on DM ($\approx 10^{28} \text{ s}$) must decay before the production of light elements in the early universe i.e. $\tau < 1 \text{ s}$ as constrained by successful Big Bang Nucleosynthesis (BBN)  \cite{BBNLimits} (for the annihilation limits see~\cite{BBNLimitsAnnihilation}).

We find four interesting scenarios: 
\begin{itemize}
\item If $M_Q \gg \LDC$ the glueball is the LCP and we take it to be the DM candidate. The dark baryon mass is independent on the glueball mass $m_\mathcal{B} \approx N_{DC} M_Q$.  The dark baryons can be made heavy and decay before BBN. If the dark baryons are long-lived a multi-component dark matter scenario will be the consequence. 
\item Within the above mass hierarchy one can find a parameter space of the theory where dark baryons are the DM candidate. Here the glueballs will have to decay before BBN via, for example, SM residual interactions. Nevertheless being still the LCPs they now provide a thermal bath of massive particles with a finite lifetime.
 \item If there are light fermions in the spectrum with $M_Q \ll \LDC$, the LCPs are dark pions, which generically will not be long-lived on cosmological time scales. Thus the pions have to decay before BBN and provide a plasma with finite lifetime as well, taking up the glueball role. Here the dark baryons are strongly coupled states that we take them to be the DM candidate with a mass of the order of $N_{DC}\LDC$.
 \item A small fraction of the above parameter space can exhibit long-lived pions when their masses are below the GeV scale. This is indeed the case of the simplest miracle \cite{Hochberg:2014kqa} where the baryons, however, are not present in the spectrum of the theory because of the choice of the underlying dynamics. If dark baryons are present one will have to consider the presence of additional stable relic states. 
 \end{itemize}

\begin{figure}[t]
\begin{center}
\includegraphics[width=0.65\textwidth]{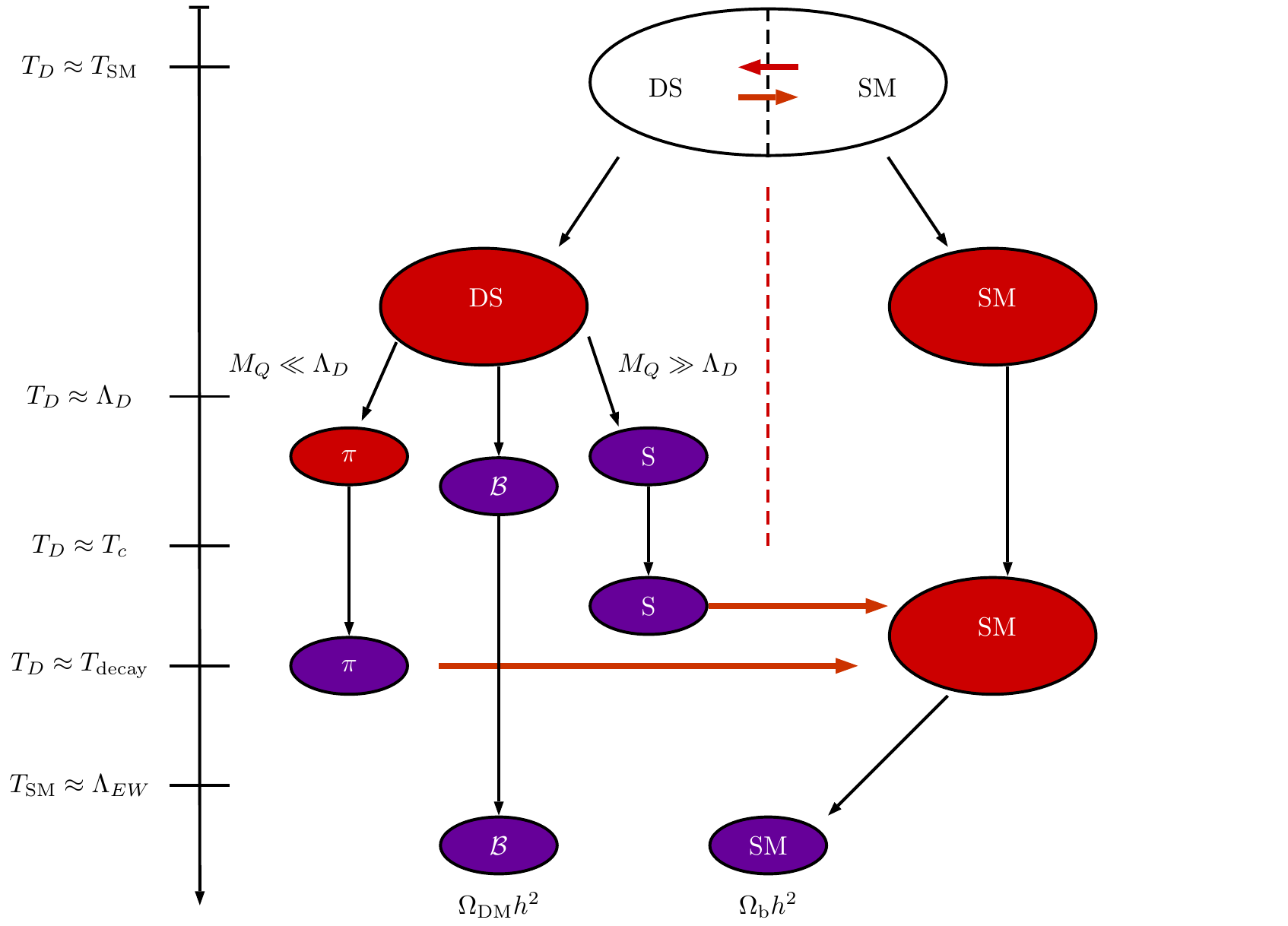}
\caption{\label{fig:cartoon}  Schematic overview of the thermal history, of a non-abelian dark sector. At high temperatures, we expect the systems to be in thermal contact. After this connection is lost, the dark force confines and forces the dark quarks in dark-color neutral bound states. Baryons ($\mathcal{B}$) and glueballs ($S$) are non-relativistic at confinement, pions ($\pi_D$) on the other hand become non-relativistic substantially later. Relativistic particles are denoted by red ovals, non-relativistic particles are represented by purple ovals. The number-changing interactions can heat up the LCP plasma, but eventually, they decouple at $T_c$ and a phase begins, where dark particle number in the plasma is conserved. Dark baryon freezeout can happen in the number-changing and number-conserving era of the dark (or LCP) plasma and the freeze-out process will be different. Finally, the decay of the dark plasma to SM final states partially re-introduces a thermal link of the systems at $T_{\rm decay}$. The decay of the LCPs can furthermore lead to entropy injection into the SM bath and dilute the dark baryon relics.}
\end{center}
\end{figure}

As sketched in Fig. \ref{fig:cartoon}, we observe that the ingredient which is common to all scenarios, is the existence of an LCP plasma made of massive particles (glueballs/pions). This plasma will generically lose thermal contact with the SM and undergo a distinct thermal history once the LCPs become non-relativistic \cite{Canibal1, Canibal2}. In particular number changing interaction among the LCPs will lead to unusual phenomena in the dark plasma. The relevant quantity we will excessively use, is the cross section, which controls the relevant number changing process. In particular for the $3 \rightarrow 2$ process, it can be found from the matrix element of the underlying process~\cite{1702.07716}
\begin{align}
\med{\sigma v^2}_{123 \rightarrow 45} = \frac{g_4\,g_5}{64 \pi S_f m_1 m_2 m_3} \times \lambda^{1/2}\left( m_1 + m_2 + m_3, m_4, m_5 \right) \overline{ |\mathcal{M}^2|}\,,
\end{align}
where $g_i$ are the number of degrees of freedom of the final state particles, $S_f$ the symmetry factor, $m_i$ the corresponding particle masses and $\lambda(x,y,z) = (1 -(z+y)^2/x^2 )(1- (z-y)^2/x^2)$. Matrix elements for the number changing processes for many symmetry breaking patterns can be found in Ref.~\cite{1411.3727}.

Finally, due to the finite lifetime of the LCPs, their decay will partially restore the thermal contact with the SM at late times. To investigate the resulting intriguing phenomenology, we will briefly review the unusual thermodynamics of a non-relativisitc dark plasma.

\section{Dark Thermodynamics} \label{sec:darkthermodynamics}
\label{sec:thermo}

In DM theories with a nontrivial spectrum with a mass gap, the freezeout will show strong deviations from the standard scenario, as discussed in \cite{Canibal1, Canibal2}. The qualitative picture one has to have in mind is the following. Dark matter freezeout happens when the dark sector has already confined, and it mainly consists of dark baryons, as they are guaranteed to be long-lived thanks to dark baryon-number conservation. The light part of the spectrum (pions or in suitable cases glueballs) are relativistic and can influence the universe expansion leading to a freeze-out process for $\mathcal{B}  +\bar{\mathcal{B}} \rightarrow \rm LCPs$ which is non-standard. Especially in the cases of heavy glueballs, however, no clear hierarchy between scales is present and they might contribute to the relic abundance. Despite the fact that LCP decay is not symmetry protected, they could be long-lived on cosmic scales, if sufficiently light. We will come back to this possibility in \ref{sec:BoltzmannNumberConserving}. Here we will summarize the conditions which lead to non-standard freezeout and that have to be satisfied before the DM (dark baryon) annihilation reactions decouple. 

\begin{itemize}
\item First of all dark sector particles must not be in kinetic equilibrium with SM particles at the freeze-out temperature, meaning that at that temperature scale the reactions $\chi \text{SM} \rightarrow \chi \text{SM}$ have to be weak. This condition is generically satisfied if the lightest composite particles are SM singlets. This is realized when the lightest dark quarks in the spectra are SM singlets and heavier dark quarks establish the thermal link at higher temperatures. We will also discuss a few exceptions to this scenario.
\item Furthermore, the lightest particle of the dark sector (here the LCPs) is stable on cosmological scales i.e. $\Gamma_{\rm LCP} \ll H(T)$.
\item The freeze-out temperature is below the LCP mass so that the LCP is non-relativisitc.
\item If the number-changing processes $3 \,\text{LCP} \rightarrow 2 \,\text{LCP}$ are strong enough to maintain chemical equilibrium at freezeout i.e. $\med{\sigma v^2}_{3 \rightarrow 2} n_{\rm LCP}^2 \gg H(T)$, the resulting hotter dark plasma leads to a non-standard freezeout. Here and in the following, we denote with $v$ the relative velocity between initial particles in a $2\rightarrow2$ process, this normalization is also used in multiparticle processes with the appropriate power for dimensional reasons. The $\med{\sigma v^2}_{3 \rightarrow 2}$ is the cross section of the $3 \rightarrow 2$ process, which we defined in the previous section.
\item If number-changing processes are inefficient at the dark freezeout, a chemical potential develops in the dark sector. This is a new phenomenon that must be taken into account for the proper determination of the DM relic abundance since the dark sector temperature has a non-standard scaling as a function of the scale factor of the universe in this case.
\end{itemize} 

The most natural realization of such a thermally secluded dark plasma arises in models, where the DM is made up of SM singlet dark quarks and the thermal link is established by a heavier bridge dark quark. A minimal set-up of this type is a singlet plus doublet construction,  where the doublet establishes the thermal link at higher temperatures. See for a complete list of viable quantum numbers Ref.~\cite{AccidentalDM}.

Thermodynamics will be extremely helpful to describe the evolution of these systems, for a summary of multi-fluid thermodynamics, see appendix~\ref{appendixB}. In particular, entropy conservation plays an important role since the dark sector entropy is dominated by the lightest dark degrees of freedom comprising the dark plasma.  The ratio of the entropy densities is fixed $\zeta = s_{\rm SM}/s_D$, and generically set by weak interactions that keep the SM and the dark sector in kinetic equilibrium at high temperatures. After those interactions decouple the entropies are separately conserved, as we assume both sectors to evolve adiabatically. While this assumption is justified for the SM, for the dark sector one has to be more careful.  For example within the scenario in which $M_Q \gg \Lambda$ the pure dark Yang-Mills theory at low energies undergoes a first-order phase transition as a function of the temperature when $N_{DC} \geq 3$.  In this case the entropy ratio established at high temperature only provides an upper bound $\zeta < s_{\rm SM}/s_D|_{T_{\rm high}}$. 

We can now differentiate two profoundly different regimes depending on the relative strength of the number-changing process.  

In the following, we will refer to the generic LCP as $\pi_D$ (keeping the subscript $D$ implicit) although the formalism applies to dark LCP glueballs as well. 

\subsection{Non Conserved Dark Particle Number}

Generically at large temperatures, the number-changing processes will maintain chemical equilibrium in the dark sector and enforce a vanishing chemical potential.  The crucial feature of this regime is that while the temperature in the SM sector scales as $T_{\rm SM} \propto 1/a $ the dark sector temperature scales as  $T_D \propto 1/\log{(a)} $, which leads to a large temperature difference between the two sectors, see for example Ref.~\cite{Canibal1}.
%
 %
 This is in strong contrast to the case where the dark sector is dominated by relativistic degrees of freedom (say a dark photon) and the temperatures are linearly dependent. We thus refer to the above possibility as the hot dark plasma phase. 

The relevant thermodynamic features follow from entropy conservation. Since the LCPs are a  non relativistic fluid with zero pressure $P=0$, we have $s_D = \rho_D/T_D \approx m_\pi n_\pi/T_D$. At the same time the SM entropy density is  $s_{\rm SM} = \,2\pi^2/45~ g_{\rm SM} T_{\rm SM}^3$. Now, from $s_{\rm SM} = \zeta~ s_D$ we get the relation 
\begin{align}
\label{eq:TempRatio}
T_{\rm SM}^3 = \left( \frac{45}{\sqrt{32 \pi^7}} \right) \left( \frac{\zeta~g_{\pi}}{g_{SM}} \right) \sqrt{m_\pi^5 T_D}\,
e^{-\frac{m_{\pi }}{ T_D}}  ~.
\end{align}
This shows that the dark sector gets significantly hotter than the SM at scales where the relevant degrees of freedom are the LCPs counted by $g_\pi$. Clearly a dramatic feature of the hot dark plasma era. 

An important quantity we will need is the Hubble rate as a function of the dark sector temperature. Solving the Friedmann equation we obtain


\be\label{eq:HubbleHot}
H(z_\pi) \approx  \left\{
\begin{array}{ll}
\left( \frac{3\sqrt{5}}{4 \pi^{5/2}} \frac{\zeta^2 g_{\pi}^2}{\sqrt{g_{SM}}} \right)^{1/3} \frac{m_{\pi}^2}{M_{Pl}}~ z_{\pi }^{-1/3} e^{-\frac{2 z_{\pi }}{3}}  & \mathrm{for} \quad (\rho_{\rm SM} > \rho_D )\,,  \\  \\
\sqrt{\frac{g_\pi \sqrt{8}}{3
   \sqrt{\pi}}} \frac{m_{\pi}^2}{M_{Pl}} ~ z_{\pi}^{-3/4} e^{-\frac{z_{\pi }}{2}} & \mathrm{for} \quad  (\rho_{D} > \rho_{\rm SM}) \,,
\end{array}
\right.
\ee
with $z_\pi = m_\pi/T_D$. 

The Hubble rate is an essential ingredient for the DM freeze-out computation, we are ultimately interested in. We now discuss what happens at temperatures at which the number-changing processes cannot maintain the chemical equilibrium in the plasma and $3 \, \text{LCP} \rightarrow 2 \,\text{LCP}$ interactions freeze out.

\subsection{Conserved Dark Particle Number}
The number-changing operators are bound to switch off when the reaction rate drops below the Hubble rate. This happens at a critical temperature of $T_c$.  Below this temperature, the dark particle number is conserved along with its entropy. 

Keeping into account the conservation of the comoving entropy for a non-relativistic pressureless gas we have the following conservation law: 

\begin{equation}
0 = \frac{d}{dt} \left( \frac{m-\mu}{T_D} n \right) + 3H \left( \frac{m-\mu}{T_D} \right) n \implies \dot{\mu} + (m - \mu) \frac{\dot{T_D}}{T_D} = 0 \ .
\end{equation}
Here $\mu$ is the chemical potential for the LCP. Solving the first order differential equation we obtain:   \begin{equation}
\mu(T_D) = m_\pi \left( 1 - \frac{T_D}{T_c} \right) = m_\pi \left( 1 - \frac{z_{\pi c}}{z_\pi} \right)\,. 
\end{equation}
where we used as a initial condition $\mu(T_c) =0$.

Furthermore, using  $\rho_D \propto T_D^{3/2}$ in the first Friedmann equation ($H^2 \propto \rho_D+\rho_{\rm SM}$) together with $H^2 \sim 1/a^3$, we verify that the dark temperature quickly drops as a function of the scale factor: $T_D \propto 1/a^2$. This scaling holds independently whether SM or dark sector dominates. 

The critical temperature, at which the number-changing processes decouple is found by solving $H(T_D) = \langle \sigma_{3 \rightarrow 2} v^2 \rangle  n_\pi^2$, which is equivalent to

\begin{align}
z_{\pi\,c }= \left\{
\begin{array}{ll}
 - 14.5 + \log{\left(  \zeta^{-1/2} g_{\rm SM}^{1/8} \, g_\pi \right)} +  \frac{1}{4} \log{ \left(  m_\pi^4 M_{\rm Pl}  \langle \sigma_{3 \rightarrow 2} v^2 \rangle\right)} - 2 \log{\left(z_{\pi c}\right) }  \quad (\rho_{\rm SM} > \rho_D) \,,  \\ \\
-17.4 + \log{\left(  g_\pi \right)} + \frac{2}{3} \log{\left( m_\pi^4  M_{\rm Pl}  \langle \sigma_{3 \rightarrow 2} v^2 \rangle \right)}  - \frac{3}{2} \log{\left(z_{\pi c}\right)}   \quad  (\rho_{D} > \rho_{\rm SM}) \,,
\end{array}
\right.
\end{align}
where we defined $z_{\pi\,c }= m_\pi/T_c$.

\begin{figure*}[t]
\centering
\subfloat[ \label{fig:scalingT} The temperature evolution of the dark sector (blue line) compared to the SM sector (red line). ]
{\includegraphics[width=0.45\textwidth]{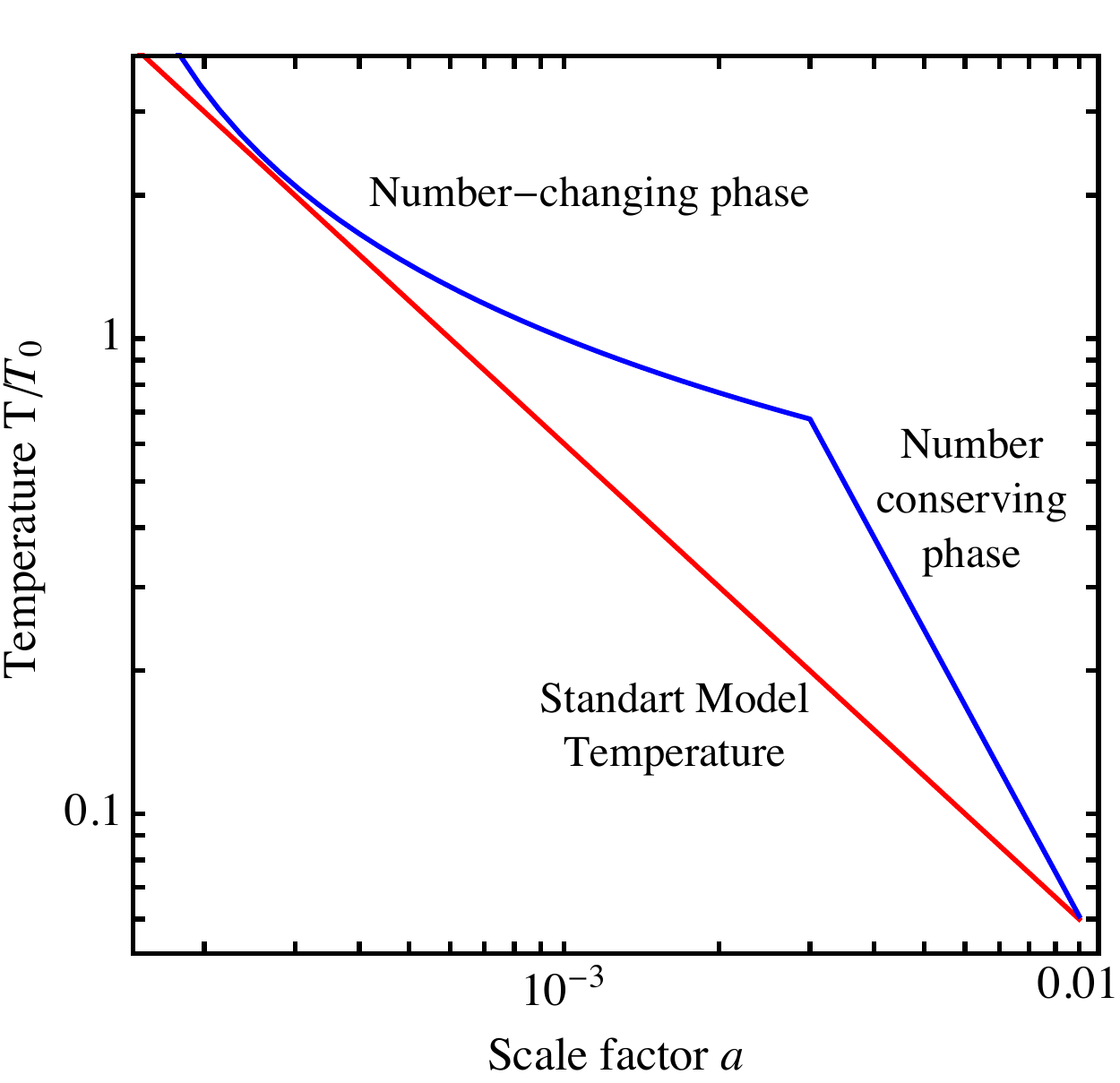}}
\hfill
\subfloat[ \label{fig:scalingrho} Red-shift of the dark sector (blue), non-relativistic matter (green) and radiation (red).]
{\includegraphics[width=0.45\textwidth]{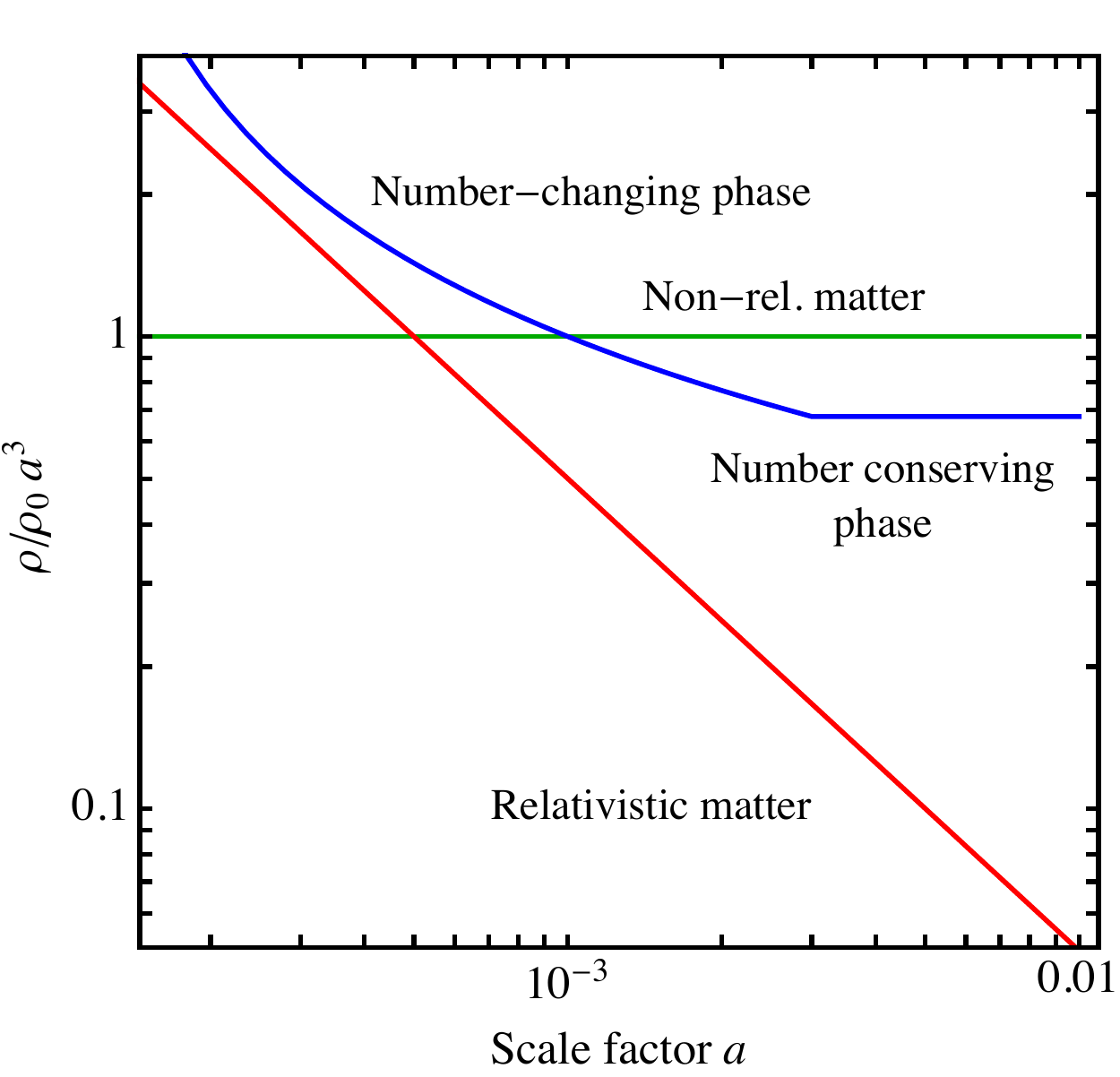}}
\caption{\label{fig:scaling} Slow temperature decrease for the dark sector in the number changing phase is due to rest mass conversion of the particles into kinetic energy (a), at the same time the energy density drops faster (b). In the number conserving phase however, the chemical potential leads to a rapid temperature decrease in the dark sector to maintain constant particle number density.}
\end{figure*}

Given that the dark sector entropy is dominated by the LCP, imposing entropy conservation $\zeta = s_{\rm SM}/s_{\pi}$ and taking into account the chemical potential one derives the new relation between the SM temperature and the dark temperature of the two fluids: 
\begin{align}
\label{eq:TempRatioChem}
T_{SM}^3= \left( \frac{45}{\sqrt{32 \pi^7}} \right) \left( \frac{\zeta\,g_\pi}{g_{SM}} \right) \sqrt{\frac{m_{\pi}^5 T_D^{3}}{T_c^2}} e^{-\frac{m_{\pi }}{T_c}} \,.
\end{align}
The interesting feature is that  now the two temperatures are no longer exponentially related. Furthermore, we have the following expression for the Hubble rate as a function of dark temperature, expressed in terms of $z_\pi$, in this regime

\be\label{eq:HubbleChem}
H(z_\pi) \approx  \left\{
\begin{array}{ll}
\left( \frac{45}{16 \pi^{5}} \right)^{1/6} \left( \frac{\zeta^2~ g_\pi^2}{\sqrt{g_{SM}}} \right)^{1/3} \left(\frac{m_{\pi}^8}{M_{Pl}^3 T_c^2}\right)^{1/3}  z_\pi^{-1}\, e^{-\frac{2 m_{\pi}}{3T_c}}  & \mathrm{for} \quad  (\rho_D < \rho_{\rm SM})\,,  \\ \\
\left(\frac{8}{9\pi}\right)^{1/4} \frac{m_\pi^2}{M_{Pl}} \sqrt{g_{\pi}}\, z_{\pi}^{-3/4} e^{-\frac{m_{\pi}}{2T_c}}  & \mathrm{for} \quad  (\rho_D > \rho_{\rm SM}) \,.
\end{array}
\right.
\ee

Finally, to determine when either of the two fluids dominates the energy budget of the universe we first note that the SM is always relativistic, as it contains mass-less particles, and thus $s_{\rm SM} = (\rho_{\rm SM} +p_{\rm SM})/T_{\rm SM}  \approx 4/3\,\rho_{\rm SM }\,T_{\rm SM}$. The temperature $T_D^{\rm e}$ (and corresponding SM temperature  $T_{SM}^{\rm e}$) for which the two energy densities are equal is defined by $\rho_{\rm SM}(T_{SM}^{\rm e}) = \rho_D(T_D^{\rm e})$. With the definition $\zeta = s_{\rm SM}/s_D$ we obtain the equation $4/3 \zeta^{-1} = T_{\rm SM}^{\rm e}/T_D^{\rm e}$. This can be expressed in terms of the dark sector inverse temperature $z_\pi^{\rm e}  = m_\pi/T_D^{\rm e}$  using eq. (\ref{eq:TempRatio}) or eq.  (\ref{eq:TempRatioChem}):
\begin{align}
\begin{array}{ll}
\frac{90}{(2\pi)^{7/2}} \frac{\zeta~ g_{\pi}}{g_{SM}} (z_\pi^e)^{5/2} e^{-z_{\pi}^e} = \left( \frac{4}{3 \zeta} \right)^3\,  & \mathrm{for}\quad ( z_{\pi c} > z_{\pi}^e ) \,,  \\ \\
\frac{90}{(2\pi)^{7/2}} \frac{\zeta~g_{\pi}}{g_{SM}} (z_\pi^e)^{3/2}~ z_{\pi c}~ e^{-z_{\pi c}} = \left( \frac{4}{3 \zeta} \right)^3\,  & \mathrm{for}  \quad ( z_{\pi c} < z_{\pi}^e ) \,,
\end{array}
\end{align}
which defines the temperature at which the dark sector starts to dominate the energy budget of the universe. In the case in which number changing LCP processes are active, this is given by the solution to:
\begin{align}
z_\pi^e \approx 5/2 \log{(z_\pi^e)} + \log{(\zeta^4 \,g_\pi/g_{\rm SM})}- 2.8\,,
\end{align}
while if the LCP number is conserved:
\begin{align}
z_\pi^e \approx 0.64 \left( \frac{e^{z_{\pi c}} g_{\rm SM}}{ g_\pi z_{\pi c} \zeta^4} \right)^{2/3} \,.
\end{align}
For all $z_\pi > z_\pi^e$ the dark sector dominates the energy budget of the universe. This will be crucially important, when possible entropy dilution of frozen-out relics will be considered~\cite{1109.2829}. 

In Fig.~\ref{fig:scaling} we show schematically the evolution of the temperature Fig.~\ref{fig:scalingT} and energy Fig.~\ref{fig:scalingrho} of the dark sector for the number changing and number conserving phases and compare them to the SM temperature and relativistic and non-relativistic matter respectively. The results of this section are summarized in appendix~\ref{appendixC}.

\section{Dark Freezeout}
\label{sec:freeze-out}

We are now ready to discuss the dark freezeout following reference~\cite{Canibal2}. One can envision the following possibilities: 
\begin{enumerate}
\item If the SM energy density dominates the universe at freezeout of $\mathcal{B} + \bar{\mathcal{B}} \rightarrow \rm LCPs$, the DM relic density is enhanced by a factor proportional to $T_D/T_{\rm SM} \gg 1$. Depending on whether the dark sector particle number is conserved or not the actual functional dependence of $T_D$ on $T_{\rm SM}$ changes.  

\item If the dark sector dominates the universe energy density at freezeout of  $\mathcal{B} + \bar{\mathcal{B}} \rightarrow \rm LCPs$, the corresponding DM relic density is enhanced by $(T_D/T_{\rm SM})^{3/2} \gg 1$. As for the point above the actual functional dependence of $T_D$ on $T_{\rm SM}$ depends on whether the dark sector particle number is conserved at freezeout. 

 \item If the LCPs dominate the universe energy budget and decay after freezeout of  $\mathcal{B} + \bar{\mathcal{B}} \rightarrow \rm LCPs$, the entropy injection dilutes dark baryon DM and suppresses it by a factor $D \propto T_{\rm SM}^e/T_{\rm RH}$, where $T_{\rm SM}^e$ is the SM temperature at which  LCPs start dominating and $T_{\rm RH} \approx 0.55~ g_{\rm SM}^* \sqrt{\Gamma_\pi M_{\rm pl}}$ is the generated reheating temperature, see section~\ref{sec:EntropyDilution} for the derivation of this analytic estimate. 
\end{enumerate}
The dark baryons annihilate in a rearrangement process.  As discussed in Ref.~\cite{WittenBaryons}, the direct annihilation process is exponentially suppressed by $e^{- c\,N_{DC}}$, in contrast to the rearrangement processes. Qualitatively, the situation can be described in the following way. Since a baryon in an $SU(N_{DC})$ theory is a $Q^{N_{DC}}$ object it carries baryon number one or equivalently fermion number $N_{DC}$. If a baryon and antibaryon collide they can rearrange into a state with $N_{DC}-1$ quark-antiquark pairs by meson emission. The remaining multiquark state has baryon and fermion number zero and thus consequently annihilates via sequential meson emission. 

The process with the maximal cross section allowed by unitarity is a geometrical interaction. The cross section is geometric when all partial waves, up to the maximally allowed value of $\ell \sim R_B \times p$ contribute. Here the baryon radius is $R_B$ and $p$ the relative momentum. In the sum over partial waves, the largest contributions come from the partial waves with large angular momenta. For this excited state to decay it needs to get rid of the angular momentum, as discussed in Ref.~\cite{ColoredRelics}. An efficient way is the emission of mesons. This explains why the process with only two final state pions, $\mathcal{B} + \bar{\mathcal{B}} \rightarrow 2 \pi_D$, has a lower rate compared to states with higher pion multiplicity. The emission of multiple pions allows a more efficient de-excitation of the higher $\ell$ states, on the other hand, energy conservation limits the number of emitted pions.  In the SM the branching ratio to three mesons dominates over the two meson annihilation for proton-anti-proton annihilation at low energies~\cite{Amsler:1997up}. Furthermore, as discussed in Ref.~\cite{hep-ex/0501020}, the five pion annihilation has the largest branching fraction, which seems to be an efficient trade-off between energy conservation and efficient decay of large angular momentum states. 

In the confining dark sector, the exact multiplicity of the final state mesons can not be predicted from first principles, however, the same logic suggests that multi-meson final states will be favored.  Adding this word of caution, we note that the exact number of mesons in the final state does not strongly affect our results. Especially, since the strongly coupled regime of the considered system does not allow to perform precise calculations, but rather order of magnitude estimates. Finally, since the dark mesons annihilate to multiple SM particles, the resulting phenomenology is a deep cascade reaction, and as shown in Ref.~\cite{1503.01773} cascades with more than a few steps lead to similar indirect detection signals. Those will be relevant for our phenomenological discussion later. For the calculation of the annihilation rate for the freezeout process the important assumption is that the cross section is geometrical, an assumption that is supported by studies of the proton anti-proton annihilation. Note that the calculation of the asymptotic DM abundance does not strongly depend on the final state pion multiplicity. We, thus, restrict ourselves to final states with $N_{DC}$ pions, since in this case helpful analytical expressions can be found. The analytical treatment of the Boltzmann equations will be sufficient for our order of magnitude calculations, more accuracy could be reached numerically, but that goes beyond the scope of our work.

In references~\cite{Canibal1, Canibal2} the sudden freeze-out approximation was adopted in order to obtain analytical results for the DM relic abundance. Here we will demonstrate that the sudden freeze-out approximation is not accurate because it does not take into account the late time annihilation processes which count here.

We start with the full Boltzmann equation for the DM baryons which reads: 
\begin{align}
\label{eq:BoltzmannGeneral}
\frac{1}{a^3}\frac{d ( a^3 n_\mathcal{B})}{dt} = -2\gamma \left( \frac{n_\mathcal{B}^2}{n_\mathcal{B}^{eq2}} -  \frac{n_\pi^{N_{DC}}}{n_\pi^{eq N_{DC}}} \right)\,.
\end{align} 
Here the space-time interaction density is $2\gamma \approx \langle \sigma_{\rm ann} v\rangle n_\mathcal{B}^{eq2}$ and further  assumes that the baryon-antibaryon annihilation takes place via a re-arrangement of the underlying degrees of freedom into $N_{DC}$ pions \cite{WittenBaryons}.  

It is convenient to rewrite the left-hand side in terms of a dimensionless quantity as 
\begin{align}
\label{eq:BoltzmannRewrite}
\frac{1}{a^3}\frac{d ( a^3 n_\mathcal{B})}{dt} = \frac{dY_\mathcal{B}}{dz} \left( \frac{dH}{dz} \right)^{-1} \left( \frac{dH}{dt}\right) s_D =  \frac{dY_\mathcal{B}}{dz} \underbrace{\left( \frac{dH}{dz} \right)^{-1} \left(- \frac{1}{m} H^2 \right) s_D}_{:= J(z)} \,,
\end{align} 
with $Y_\mathcal{B} = n_\mathcal{B}/s_D \approx n_\mathcal{B}/s_\pi$, $z = m_\mathcal{B}/T_D = z_\pi m_\mathcal{B}/m_\pi  = z_\pi/r$ and the scale factor evolves as $a \propto t^m$. The Jacobian factor in (\ref{eq:BoltzmannRewrite}) contains most of the thermodynamic information.  We now move to systematically solve the Boltzmann equations for the different cases.

\subsection{Boundary Layer Solution of the Boltzmann Equations}

A direct application of the general formula \eqref{eq:generalBE}, derived in appendix~\ref{appendixC}, gives the Boltzmann equation for dark Baryons and LCP abundances:
\begin{align}
\dot{Y}_\mathcal{B} &= - \frac{1}{2} \frac{s\, \Gamma_{1}}{J(z)} \left[ Y_\mathcal{B}^2 - \left( \frac{Y_{\pi}}{Y_{\pi}^{eq}}\right)^{w} (Y_\mathcal{B}^{eq})^2 \right], \\
\dot{Y}_{\pi} &= - w \frac{s^{w\, -1} \Gamma_1}{J(z)} \left[ Y_{\pi}^{w} - \left( \frac{Y_\mathcal{B}}{y_\mathcal{B}^{eq}} \right)^{2} (Y_\pi^{eq})^{w} \right] - \frac{s^2 \Gamma_{32}}{J(z)} \left[ Y_\pi^3 - Y_\pi^2 Y_\pi^{eq}\right].
\label{eq:baryonBE&LCPBE}
\end{align}
Moreover, we parametrized the rearrangement process of baryons into LCPs with the parameter $w$, with $w=N_{DC}$ for pions and $w=2$ for glueballs. 

The composite-model dependence enters in the quantities $\Gamma_1,\Gamma_{32}$ defined according to \eqref{eq:generalDecayW}. These interactions freezeout respectively at $z_f, z_c$, and we will assume $z_f << z_c$. This approximation allows us to study the system in terms of decoupled Boltzmann equation: the baryon freezeout when $\mu_{\pi}=0$  and the (relativistic) LCPs do not deviate from the equilibrium distribution. Afterward, the $3\rightarrow 2$ process among the LCPs can freezeout.

We now discuss boundary-layer solutions to the Boltzmann equation following \cite{Bender}. Given a general differential equation, a boundary-layer type solution is a solution that satisfies $\dot{Y}(z) \ll Y(z)$ everywhere apart from a finite number of narrow regions in which the opposite is true. These narrow regions are called boundary layers, and the size is typically controlled by a small parameter in the differential equation.  A boundary layer solution is found by matching solutions in the regions outside the layers with solutions found inside the layers.

\subsubsection{Baryon Freezeout}

The general template for the Baryon freeze-out equation is 
\begin{equation}
\dot{Y} = - \lambda f(z) \left[ Y^2 - (Y^{eq})^2  \right],
\end{equation}
where we kept a general $z$-dependent function. In the case in which the LCP is the most abundant species and drives the expansion of the universe we have the ordinary result $f(z) = A z^{-2}$, with $A$ constant for $s$-wave process only. We will find a solution with a boundary layer at $z= z_f$ to be determined. For $1/\lambda \ll 1$ and $z < z_f$ we have
\begin{equation}
Y(z) = Y^{eq}(z) - \frac{1}{2\lambda f(z)} \left( \frac{\dot{Y}}{Y} \right)_{eq} + \mathcal{O}(\lambda^{-2}).
\end{equation}

The position of the boundary layer can be estimated by finding the value of $z$ for which the LO approximation in $1/\lambda$ breaks down: 
\begin{equation}
\frac{1}{2\lambda f(z_f)} \dot{Y}_{eq}(z_f) = Y_{eq}(z_f)^2.
\end{equation}

On the other side of the Boundary layer, $z > z_f$ the inhomogeneous term can be dropped and the equation reads
\begin{equation}
\dot{Y} = - \lambda f(z) Y^2 \implies Y(z) = \frac{1}{\frac{1}{Y_{\mathcal{B}}^\infty } - \lambda \int_{z}^{\infty} dx f(x)}.
\end{equation}

These two solutions have to be matched inside the boundary layer, where can change variable to $z= z_f + \kappa Z$ with $\kappa <<1$. Defining $\mathcal{Y}(Z) = Y(z)$ the equation now becomes
\begin{equation}
\dot{\mathcal{Y}} = - \lambda \kappa f(z_f + \kappa Z) \left[ \mathcal{Y}(Z)^2 - Y_{eq}(z_f + \kappa Z)^2 \right] \implies \dot{\mathcal{Y}} = - \mathcal{Y}(Z) + \mathcal{O}(\kappa)
\end{equation}
where we used the consistent balance $\kappa = 1 / (\lambda f(z_f))$ and dropped higher orders. The solution is easily found and matched with the $z>z_f$ patch:
\begin{equation}
\mathcal{Y}(Z) = \frac{1}{D + Z} = \frac{1}{\frac{1}{Y_{\mathcal{B}}^\infty }-\lambda \int_{z_f + \kappa Z}^\infty dx f(x)} \implies D = \frac{1}{Y_{\mathcal{B}}^\infty} - \lambda \int_{z_f}^{\infty} dx f(x)
\end{equation}
as well as with the $z<z_f$ patch:
\begin{equation}
\mathcal{Y}(Z) = \frac{1}{D + Z} \sim 2 Y_{eq}(z_f)\sim \kappa + O(\kappa^2 Z) \implies D = \frac{1}{\kappa}\,.
\end{equation}
Finally, the asymptotic  relic density is calculated as:
\begin{equation}
Y_{\mathcal{B}}^\infty  = \frac{1}{\lambda} \left(f(z_f) + \int_{z_f}^{\infty} dx f(x)\right)^{-1}\,.
\end{equation}

\subsubsection{Number-changing process freezeout}

We now apply the boundary layer method to the freezeout of particles interacting with number changing operators among themselves. First let us assume that the pions are relativistic. The corresponding Boltzmann equation is
\begin{equation}
\dot{Y}_\pi = -  \left[ \frac{(1+\zeta) 2\pi^2 g_\pi M^3 }{H(M)^{1/2}} \right]^2 \Gamma_{32}\, z^{-5} \left[ Y_\pi^3 - Y_\pi^2 Y_\pi^{eq} \right]~,
\end{equation}
with the general structure
\begin{equation}
 \dot{Y}_{\pi} = - \lambda z^{-5} Y_\pi^2 (Y_\pi - Y_{eq}), \quad  Y_{eq} = \text{const.}\,
\end{equation}
Taking in the outer region $z<z_f$ (with $z\gg 1$) an ansatz of the form $Y = Y_0 + \sum_{n=1} \lambda^{-n} Y_n $ we immediatly obtain $Y_n = 0$ and $Y_0 = Y_{eq}$.  This trivially implies that the freeze out cannot happen while the pions are relativistic, consistently with the $\Gamma_{32}/H$ estimate. 

Freezeout happens when pions become non-relativistic, the general equation is then
%
\begin{equation}
\label{eq:32boltzmanngeneral}
\dot{Y} = -\lambda  f(z)[ Y^3 - Y^2 Y_{eq}]\,.
\end{equation}
In the outer region $z<z_f$ ($z\gg 1$), we proceed as in the previous section with ansatz $Y = Y_0 + \frac{1}{\lambda} Y_1$ and get
\begin{equation}
Y_1 = - \frac{1}{f(z)} \frac{\dot{Y_{eq}}}{Y_{eq}^2} = \frac{1}{f(z)} \frac{d}{dz}\left( \frac{1}{Y_{eq}}\right) \sim \frac{A}{f(z)} z^{-3/2}e^z \sim \frac{1}{f(z) Y_{eq}(z)}\,.
\end{equation}
The freeze-out temperature is determined as
\begin{equation}
Y_{eq}(z_f) \sim \frac{1}{\lambda f(z_f) Y_{eq}(z_f)} \implies Y_{eq}^2(z_f) \sim \frac{1}{\lambda f(z_f)}\,.
\end{equation}
In the post-freeze-out region the equation reduces to: 
\begin{equation}
\dot{Y} = - \lambda f(z) Y^{3} \implies Y = \frac{1}{\sqrt{ \frac{1}{\left(Y_{\pi}^\infty \right)^2} - 2\lambda \int_{z}^{\infty} f(x)dx}}~.
\end{equation}
Inside the boundary layer, choosing $\kappa = 1/(\lambda f(z_f))$ the solution reads: 
\begin{equation}
\mathcal{Y}(Z) = \frac{1}{\sqrt{D + 2 Z}} \implies D = \frac{1}{\left(Y_{\pi}^\infty \right)^2} - 2 \lambda \int_{z_f}^{\infty} dx f(x)\,.
\end{equation}
Matching in the pre-freeze-out region instead, we have 
\begin{equation}
\mathcal{Y}(Z) = \frac{1}{\sqrt{D + 2Z}} \sim 2 Y_{eq}(z_f) \sim 2 \sqrt{\kappa} + ... \implies D = \frac{1}{4\kappa}\,,
\end{equation}
leading to the following asymptotic relic abundance:
\begin{equation}
Y_{\pi}^\infty = \frac{2}{\lambda^{1/2}} \sqrt{\frac{1}{f(z_f) + 8 \int_{z_f}^{\infty} dx f(x) }}\,.
\end{equation}

Equipped with the general analytic or semi-analytic solution for the Boltzmann equations, we will study the asymptotic solutions for baryons and pions/glueballs in concrete regimes, assuming different hierarchies between $z_f$ and $z_c$. 

\subsection{Application to Freezeout in the Number-changing Era}
\label{sec:BoltzmannNumberChanging}

In the following sections, we will discuss the baryon freezeout and consider stable LCPs in section~\ref{sec:stablepions}.

First, we assume that the dark freeze-out temperature $T_f$  (due to dark baryons annihilation processes) is higher than the temperature $T_c $ at which the dark number-changing operators freezeout.  Here the number-changing processes keep the chemical potential at zero while heating up the dark plasma. Furthermore, we assume that the plasma degrees of freedom do not decay in this era i.e. $\Gamma_\pi \approx H(T_0)$ and $T_f  > T_c > T_0$.  

We  use (\ref{eq:HubbleHot}) to evaluate the Jacobian factor and the fact that the LCPs are in thermal equilibrium with vanishing chemical potential. These assumptions yield
\begin{align}
\label{eq:BoltzmannHot}
\frac{1}{ \lambda } \frac{dY_\mathcal{B}}{dz} = - f(z) \left( Y_\mathcal{B}^2 - Y_{\mathcal{B}, \, \rm eq}^{2} \right)\,.
\end{align} 
The large constant in front of the derivative ensures that asymptotic methods will give good results and is given by

\be
\lambda =  \alpha \langle \sigma_{\rm ann.} v\rangle M_{\rm Pl} m_\mathcal{B} \,,
\ee

with 

\be\label{eq:alpha}
\alpha = \left\{
\begin{array}{ll}
\frac{1}{6} \, \left( \frac{g_\pi}{12 \pi^2 \zeta^2}  \sqrt{\frac{g_{\rm SM} }{10 r}}   \right)^{1/3}   & \mathrm{for} \quad (\rho_{\rm SM} > \rho_D )\,,  \\ \\ 
\frac{ r^{1/4} \sqrt{3\, g_\pi}  }{8 \, 2^{1/4} \pi^{5/4}}   & \mathrm{for} \quad  (\rho_{D} > \rho_{\rm SM}) \,. \nonumber
\end{array}
\right.
\ee

The temperature dependent function is given by 
\be\label{eq:fofz}
f(z) = \left\{
\begin{array}{ll}
\frac{1+ 2 r z}{z^{7/6}} e^{-z r/3} & \mathrm{for} \quad (\rho_{\rm SM} > \rho_D )\,,  \\ \\
\frac{1 + 2/3 r z}{z^{3/4}} e^{-z r/2}   & \mathrm{for} \quad  (\rho_{D} > \rho_{\rm SM}) \,, \nonumber
\end{array}
\right.
\ee
with $r = m_\pi/ m_\mathcal{B}$ being the ratio of LCP and dark baryon masses.  The explicit form of the equilibrium baryon number density, normalised to dark entropy is: 
\begin{align}
\label{eq:YBeq}
Y_{\mathcal{B}, \,\rm eq} =  \frac{g_\mathcal{B} e^{(r-1) z}}{g_{\pi } r^{5/2} z}\,.
\end{align}  
We determine the freeze-out temperature expanding the solution of the Boltzmann equation in powers of $1/\lambda$. The freezeout is defined as the temperature at which the next to leading order in $1/\lambda$ is of the same size as the leading order contribution. Thus defining $Y_\mathcal{B} \approx Y_\mathcal{B}^0 + Y_\mathcal{B}^1/\lambda = Y_{\mathcal{B}, \,\rm eq}  +  Y_\mathcal{B}^1/\lambda$ we can insert this ansatz in eq. (\ref{eq:BoltzmannHot}) and solving $Y_{\mathcal{B}, \,\rm eq}  =   Y_\mathcal{B}^1/\lambda$, we obtain for the case that the SM dominates the energy budget

\be
 z_f \approx \frac{ \log \left[ \frac{g_\pi\, r^{5/3}}{g_\mathcal{B} \, \lambda} \right] +  \log \left[ \frac{z_f^{7/6} (1 -(r-1) z_f)}{ 1 + 2 r z_f  } \right] }{\left( \frac{2}{3} r - 1 \right)}\,,
\ee
and for the case when the dark sector dominates we have 

\be
 z_f \approx  \frac{ \log \left[ \frac{g_\pi^{5/6} g_{\rm SM}^{5/6}r^{9/4} }{g_\mathcal{B} \zeta^{2/3} \lambda} \right] +  \log \left[ \frac{z_f^{3/4} (1 -(r-1) z_f)}{ 1 + \frac{2}{3} r z_f  } \right] }{\left( \frac{r}{2} - 1 \right)}\,.
\ee

Following the boundary layer method suggested in \cite{Bender} we find the asymptotic solution to equation (\ref{eq:BoltzmannHot}) to be 
\begin{align}
Y_\mathcal{B}(\infty) \approx \left( \lambda f(z_f) + \int_{z_f}^{z_c} \lambda \,f(z) dz \right)^{-1} 
\end{align}

In the case of $\langle \sigma_{\rm ann. } v \rangle \approx \rm const.$ the integral can be found analytically (assuming $z_c \gg z_f$)

\be\label{eq:fofz}
\frac{\int_{z_f}^\infty  \,f(z) dz}{f(z_f)}
\approx \left\{
\begin{array}{ll} 
 \frac{6 z_f}{1 + 2 r z_f}    & \mathrm{for} \quad (\rho_{\rm SM} > \rho_D )\,,  \\ \\
 \frac{4 z_f \left(e^{\frac{r z_f}{2}} E_{\frac{3}{4}}\left(\frac{r z_f}{2}\right)+1\right)}{2 r z_f+3}  & \mathrm{for} \quad  (\rho_{D} > \rho_{\rm SM}) \,, \nonumber
\end{array}
\right.
\ee
where $E_n(x)$ is the $n-$th exponential integral.  We see at this point again that the post freeze-out regime dominates over the standard scenario where 
$\int_{z_f}^\infty  \,f(z) dz / f(z_f)  \approx z_f$.  

\subsection{Application to Freezeout in the Number-conserving Era}
\label{sec:BoltzmannNumberConserving}

Next, we consider the case $T_c > T_f $ such that freezeout occurs when number-changing processes are inactive and a chemical potential leads to conservation of dark particle number. We still assume that the plasma degrees of freedom do not decay in this era i.e.  $T_c  > T_f > T_0$.  

There will  still be two possibilities, depending on whether the SM or the dark sector dominates the universe at DM freezeout.  We  use (\ref{eq:HubbleChem}) to evaluate the Jacobian factor together with the chemical potential  $\mu = m_\pi (1 - T_D/T_c)$, which implies for $n_\pi \approx n_\pi^{eq} e^{\mu/T_D}$. The Boltzmann equation (\ref{eq:BoltzmannGeneral}) reads now 
\begin{align}
\label{eq:BoltzmannChem}
\frac{1}{ \lambda } \frac{dY_\mathcal{B}}{dz} = - \tilde{f}(z) \left( Y_\mathcal{B}^2 - \tilde{Y}_{\mathcal{B},\, \rm eq}^{2 } e^{N_{DC}\, \mu/T_D} \right)\,. 
\end{align} 
The equilibrium baryon number, normalised to dark entropy, reads
\begin{align}
\label{eq:YBeq}
\tilde{Y}_{\mathcal{B}, \, \rm eq} = \frac{g_\mathcal{B} \, e^{- z + r z_c}  }{g_{\pi } r^{5/2} z_c} \quad  \text{  with  } \quad  z_c = \frac{m_\mathcal{B}}{T_c}.
\end{align} 
The temperature dependent function reads 
\be\label{eq:fofz}
\tilde{f}(z) = \left\{
\begin{array}{ll}
\frac{3 z_c^{1/3} \, e^{-z_c r/3} }{z^{3/2}}  & \mathrm{for} \quad (\rho_{\rm SM} > \rho_D )\,,  \\ \\ 
\frac{z_c \,  e^{-z_c r/2} }{z^{7/4}}    & \mathrm{for} \quad  (\rho_{D} > \rho_{\rm SM}) \,. \nonumber
\end{array}
\right.
\ee
Following the procedure outlined in the previous subsection the freeze-out temperature is 
\be
 z_f \approx \frac{ \log \left[ \frac{g_\pi (2 - r N_{DC}) r^{8/3} }{12 g_\mathcal{B} \, \lambda z_c^{1/3}} \right]  + z_c r \left(\frac{N_{DC}}{2} +\frac{1}{3} \right)+  \frac{5}{2} \log \left[z_f\right] }{\left(  (\frac{N_{DC}}{2} +1 ) r - 1 \right)}\,,
\ee
in the regime where the SM dominates the energy budget. The case in which the dark sector dominates the energy density gives
\be
 z_f \approx \frac{ \log \left[ \frac{g_\pi^{5/6} g_{\rm SM}^{5/6} (2 - r N_{DC}) r^{9/4}  z_f^{11/4} }{ g_\mathcal{B} \, \lambda z_c \zeta^{2/3} } \right]  + z_c r \left(\frac{N_{DC}}{2} + 1 \right) -2.1 }{\left(  (\frac{N_{DC}}{2} + 2 ) r - 2 \right)}\,. 
\ee
The boundary layer method leads to the following asymptotic solution for  (\ref{eq:BoltzmannChem})
\begin{align}
Y_\mathcal{B}(\infty) \approx \left( \lambda \tilde{f}(z_f) + \int_{z_f}^\infty \lambda \,\tilde{f}(z) dz \right)^{-1}  \ .
\end{align}
In the case of $\langle \sigma_{\rm ann. } v \rangle \approx \rm const.$ the integral can be solved analytically (assuming $z_c \ll z_f$)
\be\label{eq:fofz}
\frac{\int_{z_f}^\infty  \,\tilde{f}(z) dz}{\tilde{f}(z_f)}
\approx \left\{
\begin{array}{ll}  
 2 z_f  & \mathrm{for} \quad (\rho_{\rm SM} > \rho_D )\,,  \\ \\
\frac{4}{3} z_f & \mathrm{for} \quad  (\rho_{D} > \rho_{\rm SM}) \,. \nonumber
\end{array}
\right.
\ee
We find that also in this scenario, the late time annihilations play a significant role and corrections to the sudden freeze-out approximation are substantial. These solutions are applicable if freezeout happens on timescales on which the LCPs are stable. 

\subsection{Application to Freezeout with Decaying LCPs}
\label{sec:BoltzmannLCPDecay}

Here we consider the decay of the LCPs and consider two scenarios, with different effects on the asymptotic dark baryon abundance. 
\subsubsection{LCP Decay before Baryon Decoupling}

Here we assume that the lifetime of the universe becomes larger than the LCP lifetime, while dark baryons are still in chemical equilibrium with the LCP plasma. In this case the set of Boltzmann equations governing the system reads
\begin{align}
& \frac{dY_\mathcal{B}}{dt} = -2 \gamma \left(  \frac{Y_\mathcal{B}^2}{Y_{\mathcal{B}, \, \rm eq}^{2}} - \left( \frac{Y_\pi}{Y_{\pi \, \rm eq}} \right)^{N_{DC}} \right) \label{eq:BaryonBoltz}\\
& \label{eq:LCPBoltz} \frac{dY_\pi}{dt} = 2 \gamma \left(  \frac{Y_\mathcal{B}^2}{Y_{\mathcal{B} \, \rm eq}^{2} } - \left( \frac{Y_\pi}{Y_{\pi \, \rm eq} } \right)^{N_{DC}} \right) - 3 \gamma_{32}\left(  \frac{Y_\pi^3}{Y_{\pi\,\rm eq}^{3}} - \frac{Y_\pi^2}{Y_{\pi \,\rm eq}^{2}}\right)  - \gamma_{\rm ann} \left(   \frac{Y_\pi}{Y_{\pi \, \rm eq} }  -1 \right) \,,
\end{align}
where the space time interaction densities are given by $2 \gamma \approx \langle \sigma v \rangle n_{\mathcal{B} \, \rm eq }^{2} $ and $\gamma_{\rm ann} \approx \Gamma_{\rm ann} n_{\pi \, \rm eq} $ and $ 3\gamma_{32} \approx \med{\sigma_{3 \rightarrow 2} v^2} n_{\pi \, \rm eq}^{3}$. Here we allow for annihilation rate of the LCPs into SM particles. 

The LCP plasma begins to decay away once the Hubble rate drops below the annihilation rate $H(T_{\rm Decay}) < \Gamma_{\rm ann}$, which defines the decay temperature $T_{\rm Decay}$. The dimensionless number densities are now conveniently defined with respect to the SM entropy density $Y = n/s_{\rm SM}$ during the radiation dominated era. 

As discussed in \cite{CosmoDM}, once the annihilation rate is faster than the Hubble rate, the left hand side of (\ref{eq:LCPBoltz}) can be neglected at leading order and the equation becomes algebraic. Now, the effective equation for the DM number density can be obtained by inserting the solution of  (\ref{eq:LCPBoltz}) into (\ref{eq:BaryonBoltz}). We obtain
\begin{align}
& \frac{dY_\mathcal{B}}{dt} = \frac{d Y_\mathcal{B}}{dz} H(z) z s_{\rm SM}= -2 \gamma_{\rm eff} \left(  \frac{Y_\mathcal{B}^2}{Y_{\mathcal{B}\, \rm eq}^{2}} - 1  \right)\,,
\end{align}
where the effective interaction density is for $N_{DC} = 3$
\begin{align}
& 2 \gamma_{\rm eff}  \approx 2 \gamma  \frac{\gamma_{\rm ann} \,  \left( 1 + \frac{ N_{DC} \gamma_{\rm ann}  2 \gamma  3\gamma_{32}}{(\gamma_{\rm ann} +  N_{DC}\, 2 \gamma )^3} \right)  }{\gamma_{\rm ann} + N_{DC} \, 2\gamma } \approx \med{\sigma v} n_{\mathcal{B}\, \rm eq}^{2}  BR(z) \,,
\end{align}
with the branching ratio defined as 
\begin{align}
BR(z) =  \med{\sigma v} n_{\mathcal{B} \, \rm eq}^{2} \left(1 +\frac{N_{DC} \Gamma_{\rm ann}  n_{\mathcal{B} \, \rm eq}^{2}  n_{\pi \, \rm eq}^{4} \med{\sigma v}   \med{\sigma_{3 \rightarrow 2} v^2} }{\left(  \Gamma_{\rm ann} n_{\pi \, \rm eq}  +  N_{DC} \med{\sigma v} n_{\mathcal{B} \, \rm eq}^{2}  \right)^3}\right)  \left(   1 + \frac{ N_{DC} \med{\sigma v} n_{\mathcal{B} \, \rm eq}^{2}}{\Gamma_{\rm ann} n_{\pi \, \rm eq} } \right)^{-1}\,.
\end{align}
In fact baryons in large color groups can be constructed by means of the two index antisymmetric representation~\cite{3quarkBaryons}, in which case this formulae apply to arbitrary $N_{DC}$.

This final Boltzmann equation can be integrated and yields 
\begin{align}
Y_\mathcal{B}(\infty) \approx \left(\frac{ \lambda \med{\sigma v}  BR(z_f)}{z_f^2}  + \int_{z_f}^\infty \frac{ \lambda \,\med{\sigma v}  BR(z)}{z^2} dz \right)^{-1} \,. 
\end{align}
This effective description clearly shows  how the decay of the LCP generates a thermal link between the SM and the dark sector. The crucial factor is the interaction proportional to $\Gamma_{\rm ann} \left(   Y_\pi  - Y_{\pi \, \rm eq}  \right) $ in the Boltzmann equation, which establishes the thermal link, provided that $\Gamma_{\rm ann} \gg H(T_{\rm Decay})$. Below the temperature $T_{\rm Decay}$, the LCPs can not be regarded as stable any longer, as the Hubble time starts exceeding their life-times.

\subsubsection{LCP Decay after Baryon Decoupling}
\label{sec:EntropyDilution}

If the dark baryons have already decoupled from the dark plasma at the time of LCP decay, the only effect which can be relevant is entropy injection and DM dilution~\cite{0712.1031, 1805.01473, 1811.06975}. 
This is relevant if, at the time of the LCP decay, they are the dominant energy component of the universe. The condition for the dominance of LCP energy density has been 
discussed in the previous sections. After the LCPs decay to SM particles the relic abundance of the dark baryons will be diluted by
\begin{align}
\Delta = \frac{s_{\rm SM}^{\rm rh}}{s_D^{\rm Decay}}\,.
\end{align}
To find this ratio four facts are crucial:
\begin{enumerate}
\item Before the LCP decay there is a point in the evolution of the universe where $\rho_D^e = \rho_{\rm SM}^e$.
\item At the decay of the LCPs their entire energy is injected in the SM sector and reheats it $\rho_D^{\rm Decay} = \rho_{\rm SM}^{\rm r.h.}$.
\item We can equate ratios of covariantly conserved quantities, since they are just related by the ratio of the initial and final volume of the universe i.e. we have $\rho_D^{e}/\rho_D^{\rm Decay}=s_D^{e}/s_D^{\rm Decay}$.
\item Before the decay of the LCPs the SM and dark entropy ratio remains constant, in particular we use $s_{\rm SM}^e/s_{D}^e = \zeta$. 
\end{enumerate}
We therefore, have:
\begin{align}
\rho_{\rm SM}^e = \rho_D^e = \rho_D^{\rm Decay} \frac{s_D^e}{s_D^{\rm Decay}} =  \rho_D^{\rm Decay}  \frac{s_D^e \, s_{\rm SM}^e }{s_{\rm SM}^e  \, s_D^{\rm Decay} } =  \rho_D^{\rm Decay}  \frac{ s_{\rm SM}^e }{ \zeta \,  s_D^{\rm Decay} }  \, ,
\end{align}
and can rewrite:
\begin{align}
\Delta = \frac{s_{\rm SM}^{\rm r.h.}}{s_D^{\rm Decay}} = \frac{s_{\rm SM}^{\rm r.h.} \, \rho_{\rm SM}^e \, \zeta}{\rho_{\rm SM}^{\rm r.h.} \,  s_{\rm SM}^E} = \frac{T_{ \rm SM }^e \,  \zeta }{ T_{ \rm SM }^{ \rm r.h.}} \, . 
\end{align}
The reheating temperature is determined by the condition that the LCP lifetime equals the Hubble time $\tau_H = \Gamma_{\rm ann}^{-1}$. Which for a matter dominated universe is $\tau_H = 2/3 H[\rho_D^{\rm Decay}]^{-1}$. We make use of the fact that all of the dark sector energy density is converted to SM energy density in the reheating process and  rewrite the Friedmann equation 
\begin{equation}
H[ \rho_D^{\rm Decay} ]^2 = \frac{8 \pi}{3 M_{\rm Pl}^2 } \rho_D^{\rm Decay}\, ,
\end{equation}
to find the reheating energy density
\begin{align}
\rho_D^{\rm Decay}  = \frac{\left( \Gamma_{\rm ann} M_{\rm Pl} \right)^2 }{6 \pi } =  \rho_{\rm SM}^{\rm r.h.}\, . 
\end{align}
The reheating temperature is therefore given by 
\begin{align}
T^{\rm r.h.}_{\rm SM}  = \left( \frac{5}{ \pi^3 g_{\rm SM} } \right)^{1/4} \left(  M_{\rm Pl}  \Gamma_{\rm ann}\right)^{1/2}\,.
\end{align}
The dark baryon relic abundance is then given by $Y_{\rm B} =Y_{\rm B}(\infty )/\Delta$. In the case that the LCP energy is a subdominant fraction of the total energy density in the universe, no dilution takes place.

The questions regarding whether the dark baryon freezeout takes place during the number-changing or preserving era, as well as what is the actual LCP lifetime can be answered in a concrete model. Note that the number-changing interactions are generically present among LCPs (glueballs or pions) at low energies. We will study the most economical and appealing models of this type in the following sections.


\section{Weak and strong coupled regimes of dark bound states } 
\label{sec:models}

Here we employ the formalism developed in the previous sections to investigate the asymptotic abundance of stable relics in different regimes of non-abelian gauge theories.
First, we focus on the strong coupling regime, where the gauge coupling at typical momenta inside the bound states is non-perturbative.


\subsection{The Strongly Coupled Bound-state Limit } \label{sec:strongcoupling}

Here we assume the vector-like fermion masses to be small compared to the dark confining scale. In this limit, the dark bound states are relativistic and interact non-perturbatively among each other.  This situation mirrors ordinary QCD.

\subsubsection{DM Self Annihilation}
Because of the QCD resemblance we can use it as an analog computer to determine DM self annihilation by properly rescaling the measured QCD proton-proton annihilation cross section.   The annihilation is exothermic since  $\sigma v =  \rm constant $ and  the $s$-wave contribution dominates.  Thus the dark baryon annihilation cross section is \cite{AccidentalDM}
\begin{align}
\sigma v_{\rm rel.} = \frac{4 \pi}{\Lambda_D^2} \approx  \frac{4 \pi}{\MDM^2} \,. 
\end{align}

\subsubsection{Dark Pion Interactions}
In the current setup the lightest dark particles are goldstone bosons described by the chiral Lagrangian, see for details appendix~\ref{appendixA}. Because the coset space of the spontaneously broken global symmetry has a non-trivial fifth homotopy group we can write, for  $SU(N_f)_R \otimes SU(N_f)_R$ broken to $SU(N_f)_V$ for $N_f>2$ the following term induced by the WZW \cite{WZW,Duan:2000dy} action:

\begin{align}
& \mathcal{L} \supset \frac{ N_{DC}}{240 \pi^2 f_\pi^5 } \varepsilon^{\mu \nu \alpha \beta} \text{Tr} \left[ \Pi \partial_\mu \Pi \partial_\nu \Pi \partial_\alpha \Pi \partial_\beta \Pi \right]  \,, \nonumber \\
& \text{ with }  4\pi \, f_\pi  \approx  \LDC \text{ and }\Pi = \sum_a \lambda_a \pi^a\, ,
\end{align}
 where  $\lambda_a$ are the broken generators. The resulting cross section for the $3 \rightarrow 2$ processes is 
 \begin{align}
 \med{\sigma_{32} v^2} =  \frac{5 \sqrt{5}   }{2^{11} \pi^5 f_\pi^{5}\, z_\pi^2}  \left( \frac{m_\pi}{f_\pi }\right)^5 \frac{N_{DC}^2}{N_\pi^3} \, \underbrace{\frac{1}{5!} \sum T^2_{\lbrace ijklm\rbrace }}_\text{= $t^2$}\,,
 \end{align}
where $T_{ijklm} = \text{Tr}[\lambda_i \lambda_j \lambda_k \lambda_l \lambda_m]$ and $\{...\}$ denotes an ordering procedure, for example $\{1,5,4,3,2\} = 1,2,3,4,5$ (Also, changed A $\rightarrow \Pi$ in equation above).
For a breaking pattern of the form $SU(N_f) \otimes SU(N_f) \rightarrow SU(N_f)$ we have  (see \cite{Hochberg:2014kqa}) $N_\pi= N_f^2-1$ and $t^2 = 4/3\, (N_f^2-1)(N_f^2-4)$. Within this framework of chiral perturbation theory the dark baryon abundance and the effects of the dark plasma can be computed. 

\begin{figure*}[t]
\centering
\includegraphics[width=0.45\textwidth]{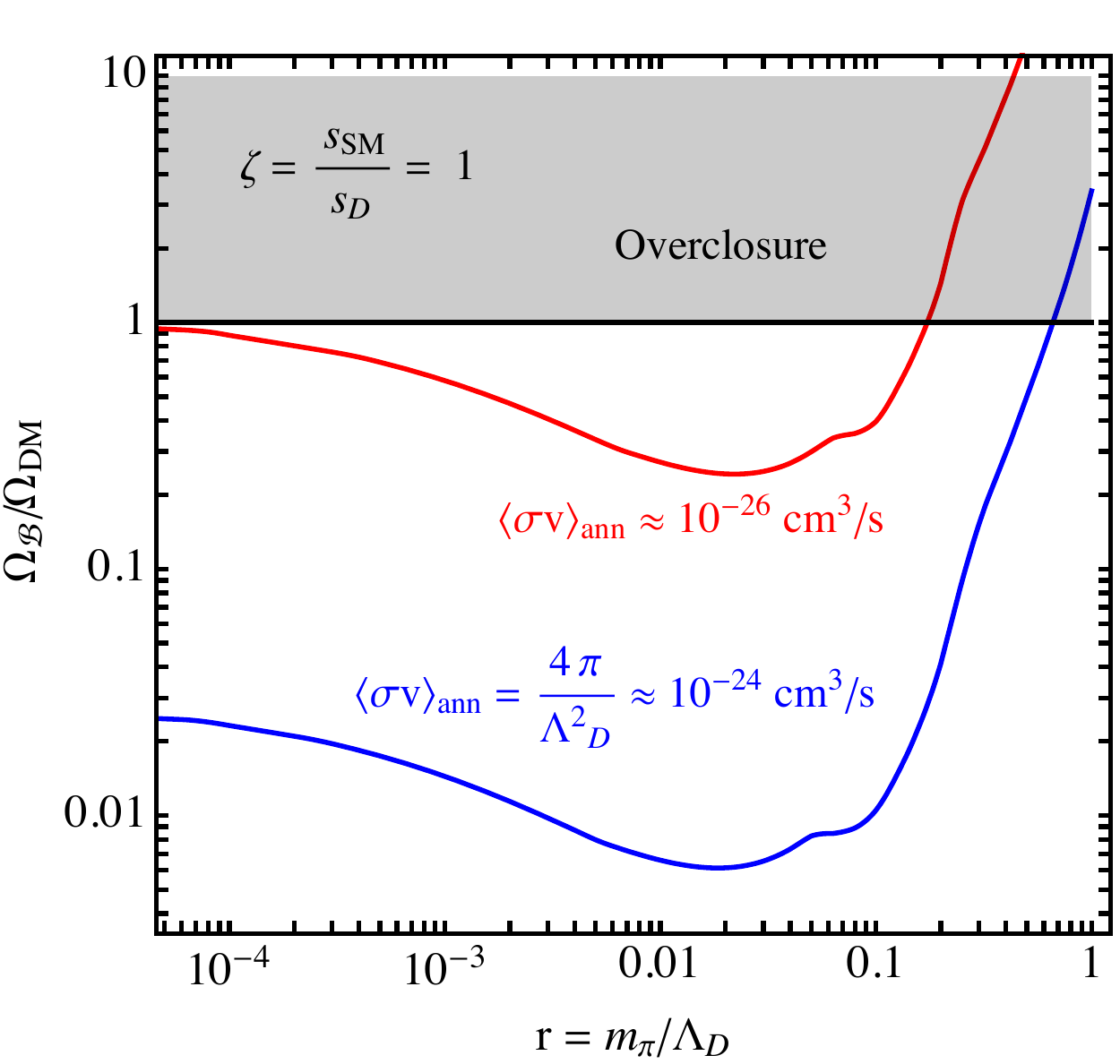}
\caption{\label{fig:RelicDCartoon} The predicted dark baryon relic density as a function of the dark pion to dark baryon mass ratio $r$. Two dark baryon masses of $m_\mathcal{B} \approx 10 \text{ TeV}$ leading to $10^{-24} \text{ cm}^3/\text{s}$ and $m_\mathcal{B} \approx 100 \text{ TeV}$ leading to $10^{-26} \text{ cm}^3/\text{s}$ are shown. In the second case there exist two pion to baryon mass ratios, that lead to the correct dark baryon relic density. This effect results from the fact, that at intermediate values of $r$ the relic density is suppressed with respect to the naive estimated value, while on the other hand an enhancement takes place when $r$ approaches values of order one. This enhancement leads to significantly larger target cross sections for heavy DM candidates. At $r$ values below $10^{-4}$ the pions are relativistic throughout the entire annihilation history of the dark baryons and no effect is present.}
\end{figure*}

In Fig.~\ref{fig:RelicDCartoon} we show the relic abundance for two selected dark baryon masses, resulting in annihilation cross sections of $10^{-24} \text{ cm}^3/\text{s}$ and  $10^{-26} \text{ cm}^3/\text{s}$. Even for the cross section that is significantly larger than the naively expected target cross section of $\mathcal{O}(1) \times 10^{-26} \text{ cm}^3/\text{s}$ the correct dark relic abundance can be achieved. This effect takes place, as the self-heating of the dark plasma enhances the dark baryon relic density when $r \approx \mathcal{O}(1)$. At intermediate values, the freezeout happens in the presence of the chemical potential in the dark sector and suppression of the abundance is achieved. This behavior leads to multiple solutions with the correct asymptotic DM relic density, as can be seen from the red curve in Fig.~\ref{fig:RelicDCartoon}. 

At small relative pion masses, the effects do not impact the DM relic abundance. This can be understood in the following way: the asymptotic solution of the Boltzmann equation is dominated by the inverse of the integral $\int_{z_f}^\infty f(z) \med{\sigma v}_{\rm ann} dz$, where $f(z)$ falls off with the power between $z^{-1}$ and $z^{-2}$ (as discussed in the previous sections). Numerically this leads to complete suppression of effects at $z > 10^4$. The pions become non-relativistic at temperatures of the order of $z \approx r^{-1}$. Thus, it is obvious that at $r>10^{-4}$ all the non-standard thermodynamic effects subside. 

\begin{figure*}[t]
\centering
\subfloat[ \label{fig:RelicD} No entropy dilution due to pion decay. Short lived or stable pions are assumed.]
{\includegraphics[width=0.45\textwidth]{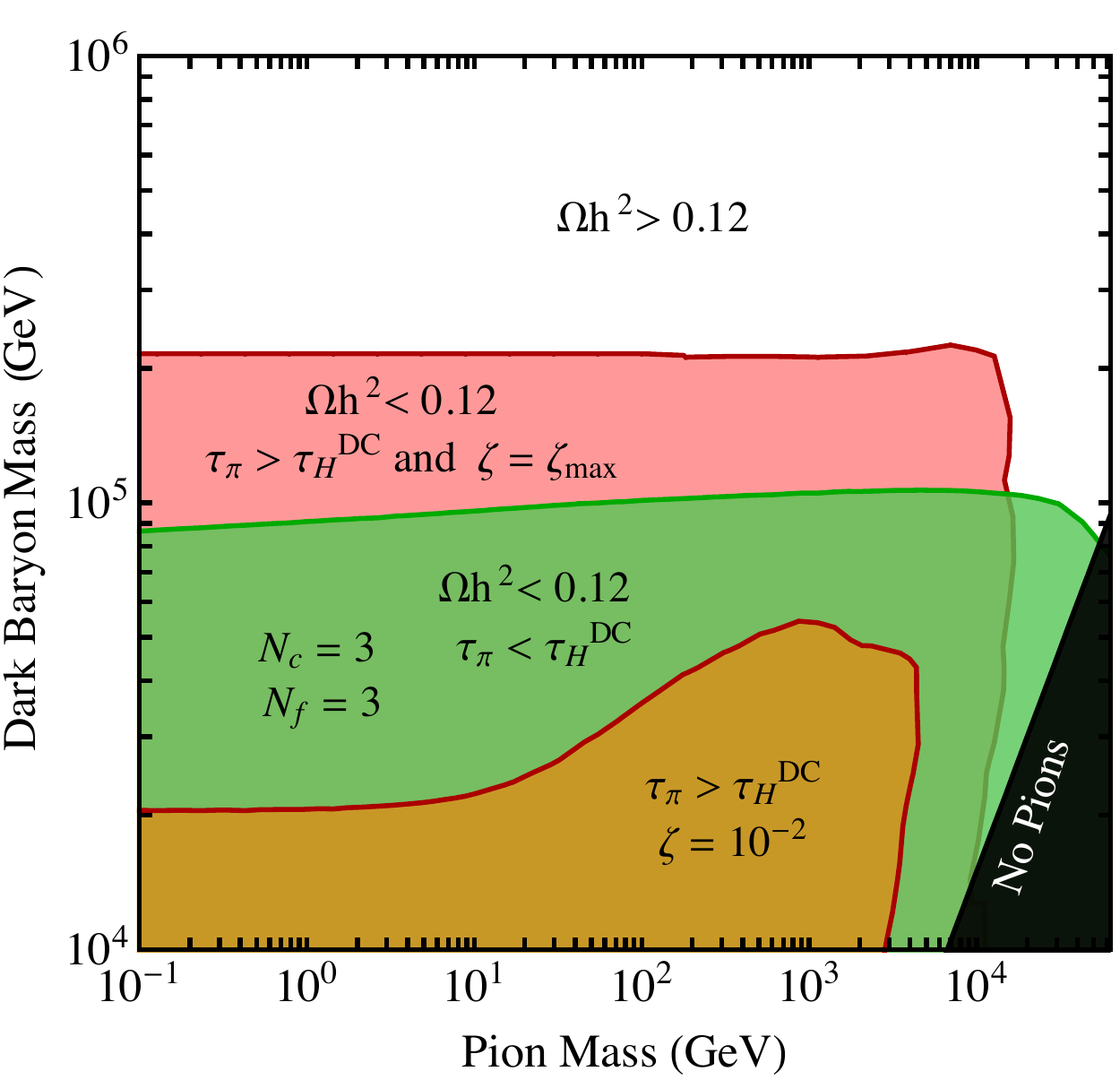}}
\hfill
\subfloat[ \label{fig:RelicDwithD}Entropy dilution in the case that pions are unstable and their lifetime is $\tau_\pi \approx 1\, \text{s}$.]
{\includegraphics[width=0.45\textwidth]{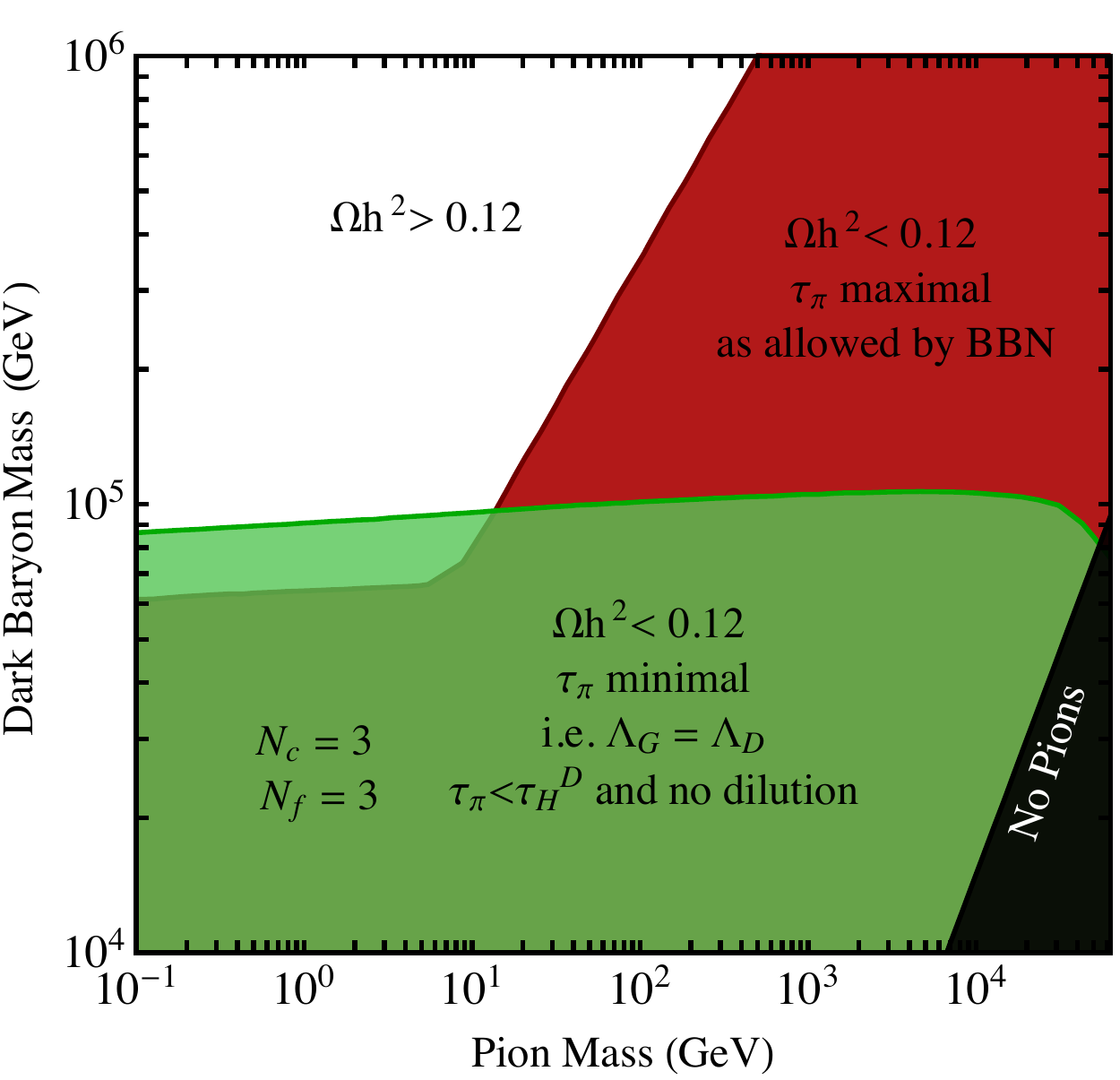}}
\caption{The model parameter space allowed by the relic density constraint in the case of $\LDC \gg M_Q$. So far it is assumed that no entropy dilution is happening. The model considered has $N_{DC} = 3$ and $N_f =3$. The light green region shows how ignoring the thermodynamic effects leads to significantly different relic abundance predictions. Note that in the strongly coupled model, the freezeout always happens in the regime with conserved dark particle number and an active chemical potential.}
\end{figure*}

Figure~\ref{fig:RelicD} shows the dark baryon relic density in the strongly coupled regime as a function of the dark baryon and pion masses while neglecting possible dilution due to pion decay. Depending on the parameter space the resulting dark baryon density will come from either number preserving or number-changing type of interactions outlined above. Therefore, the dynamics of the system is either governed by the Boltzmann equations from Section~\ref{sec:BoltzmannNumberChanging} or Section~\ref{sec:BoltzmannNumberConserving}. 
The pion lifetime is varied as a model independent free parameter and for comparison the two regimes are contrasted. In the case that the pions are stable on cosmic time-scales during the confinement and freeze-out process ($ \tau_\pi > \tau_H^{\rm DC}$) the relic abundance depends on the initial entropy ratio $\zeta$. If the pion decay is significant during the freezeout ($\tau_\pi < \tau_H^{\rm DC} $) the established connection to the SM plasma erases the $\zeta$ dependence. The Boltzmann equation used for the relic abundance is in this case described in Section~\ref{sec:BoltzmannLCPDecay}.  

 \subsubsection{Pion Decay Effects}
 
 To discuss the pion lifetime effects more concretely, we specify a model, by choosing it to contain a dark quark with SM quantum numbers. The simplest viable model, as discussed in~\cite{BaryonDM} contains only one quark multiplet with $m_Q < \Lambda_D$ is the $Q = (3, 1, 3,0)$ of $SU(3)_D \times SU(3)_c \times SU(2)_L \times U(1)_Y$.  Here the dark diagonal flavor symmetry is $SU(3)$ under which we have 8 Goldstones that decompose under  $SU(2)_L$ SM    as a $3 \oplus 5$. 
 
The dark Goldstones enjoy a G-parity symmetry (charge-conjugation followed by a weak isospin flip) that would stabilize $\pi_0$. However, the symmetry can be broken by a dimension five operator induced by physics at the scale  $\Lambda_{G \! \!  \! \!/}$. This physics may induce, for example, the operator $ - \kappa \, (\alpha_{\text{EM}}/ \Lambda_{G \! \!  \! \!/} )\, \pi_0F_{\mu \nu } \tilde{F}^{\mu \nu }$ enabling the decay with
\begin{align}
\Gamma_{\pi_0 \rightarrow \gamma  \gamma } = \frac{\kappa^2 \alpha_\text{EM}^2}{ 64 \pi^3} \frac{m_{\pi_0}^3}{\Lambda_{G \! \!  \! \!/}^2 } \,. 
\end{align}

The coefficient $\kappa$ is $\mathcal{O}(1)$ and $ \Lambda_D < \Lambda_{G \! \!  \! \!/} < M_{Pl}$, which defines the minimal and maximal lifetime of the pions respectively.  

A simple way to UV complete this decay scenario is the introduction of heavy vector-like dark quarks, which transform as $Q_2 = (3, 1, 2,1/2)$. Those will allow a Yukawa mixing with the dark quark which comprises the DM parametrized by $\epsilon$ and induces an anomaly leading to the above decay process. The decay rate, in this case, would have an additional suppression factor of $\epsilon^4$ and would thus favor longer pion life-times making the pions stable during the dark baryon freezeout. Note that this UV completion is also valid for SM singlet DM component quarks, as a Yukawa interaction is also allowed in this case~\cite{AccidentalDM}. 

It is obvious, that in order for a pion to be stable on cosmological scales its mass has to be below the GeV scale. Thus, we have a severely limited parameter space for a long-lived pion as a DM candidate. In the generic case, as mentioned before, the baryon will be the DM candidate and thus the pions have to decay before BBN i.e. $\tau_\pi <  1 \text{s}$. In the case that the dark pions have such long lifetimes, the entropy dilution, as discussed in Section~\ref{sec:EntropyDilution}, takes place. The resulting relic abundance is shown in Fig.~\ref{fig:RelicDwithD}. Note that in this case, the dark baryon mass can be significantly higher than expected from unitarity considerations~\cite{UniBound, GriestKamionkowski} or even tighter mass bounds, derived in Ref.~\cite{1911.00012}. 

\subsubsection{Stable Pion Abundance}
\label{sec:stablepions}

As discussed in the introduction, it is possible to construct models with global symmetries, in which the pions are long-lived DM candidates. This implies that the pion cannot be heavier than a few GeV. In this section, we predict the relic abundance of the long-lived pions as a function of the model parameters. 

While for the dark baryons, the freezeout and production depend to a lesser degree on the couplings of the constituent quarks to the SM particles, the situation is different in the pion case.  This is due to the fact that viable models feature light pions and therefore, depending on the concrete realization of the model, the pion-SM elastic scattering might be strong enough to maintain kinetic equilibrium between the sectors, leading to the equality of $T_D = T_{\rm SM}$.  Whether temperature equality is achieved changes dramatically the resulting freeze-out phenomenology. 

\subsubsection{Kinetic Equilibrium}

In Section~\ref{sec:darkthermodynamics}, it is shown that a dark pion sector, which is not in kinetic equilibrium with the SM bath cools slower than normal non-relativistic matter. Thus, structure formation excludes a secluded, self-heating pion sector as the sole component of the DM fluid in the universe~\cite{LaixScherrer}. 

Therefore, we first consider a scenario in which the temperature equality between the dark pion and the SM sector is maintained throughout the freeze-out process, as in~\cite{1402.5143}. Here the general Boltzmann equation Eq.~(\ref{eq:32boltzmanngeneral}) greatly simplifies and the temperature-dependent function is just $f(z) = z_\pi^{-n}$. The WZW interaction cross section includes the $z_\pi^{-2}$ scaling of the pion scattering cross section for $n=7$.  

The asymptotic number density with $n=7$ is thus given by 
\begin{align}
Y_\pi^\infty = \frac{n_\pi^\infty}{s_{\rm SM}} \approx \frac{2}{\sqrt{\lambda}} \left(  \frac{1}{(z^\pi_c)^n}+ \frac{8}{n-1} \frac{1}{(z^\pi_c)^{n-1}} \right)^{-1/2} \approx \sqrt{\frac{3}{\lambda}} (z^\pi_c)^3\,,
\end{align}
where 
\begin{align}
\lambda =  \frac{2 g_\pi^2 (1 + \zeta )^2 \pi^{5/2}}{135\, \sqrt{5} g_{\rm SM}^{1/2}}  \,m_\pi^4 M_{\rm Pl} \sigma_{32}^0\ , 
\end{align}
is the large parameter enabling the boundary layer method to work. Here we have defined 
 $\sigma_{32}^0 =  \med{\sigma_{32} v^2} z^2$ and assumed $\zeta \ll 1$.

In the case of $3 \rightarrow 2$ freezeout and equal temperatures in the SM and dark sectors, the critical temperature at which the number-changing interactions decouple can be found analytically to be 
\begin{align}
z_\pi^c = \frac{1}{2} (n-3) W\left(\frac{ \left(\frac{ 4 \pi^2 \,2^{4-n}}{45\, \lambda }\right)^{\frac{1}{3-n}}}{\pi ^{\frac{1}{3-n}} (n-3)}\right) \,,
\end{align}
where $W(x)$ is the ProductLog function. 

Since the relic density is given by $\Omega_\pi \approx  s^0_{\rm SM} m_\pi Y_\pi^\infty/\rho_c \approx 2.6 \times 10^3  \left((N_f^2-1) m_\pi  \sqrt{ \sigma_{32}^0} \, \text{ GeV}^{3/2} \right)^{-1}$ this scenario  alone does not point to a fixed target cross section, as in the case of the well known thermal WIMP \cite{WIMP}. Fixing  the relic density to be the maximally allowed relic abundance of DM we   obtain the constraint $(N_f^2-1)^2\,m_\pi^2 \sigma_{32}^0 > 10^8 \text{ GeV}^{-3}$. 

Given the scaling of  $\sigma_{32}^0 \approx 2.3\times 10^{6}\,N_{DC}^2 r^{10}\, m_\pi^{-5} N_f^{-1}$, we find that the relic density is  $\Omega_\pi \approx \mathcal{O}(1)\, r^{-5} N_{DC}^{-1} \,(m_\pi /(N_f\, \text{ GeV}))^{3/2}$ and $m_\pi \leq 0.3 \, r^{10/3} \, N_{DC}^{2/3}\, N_f  \text{ GeV}$.  Since at values of $r \approx \mathcal{O}(1)$ the chiral perturbation theory used to set up the pion Lagrangian breaks down, we find that the mass ordering scenario of $m_Q \ll \Lambda_D$ provides an upper mass bound for the pion dark matter of the order of GeV. 

\begin{figure}[t]
\begin{center}
\includegraphics[width=0.45\textwidth]{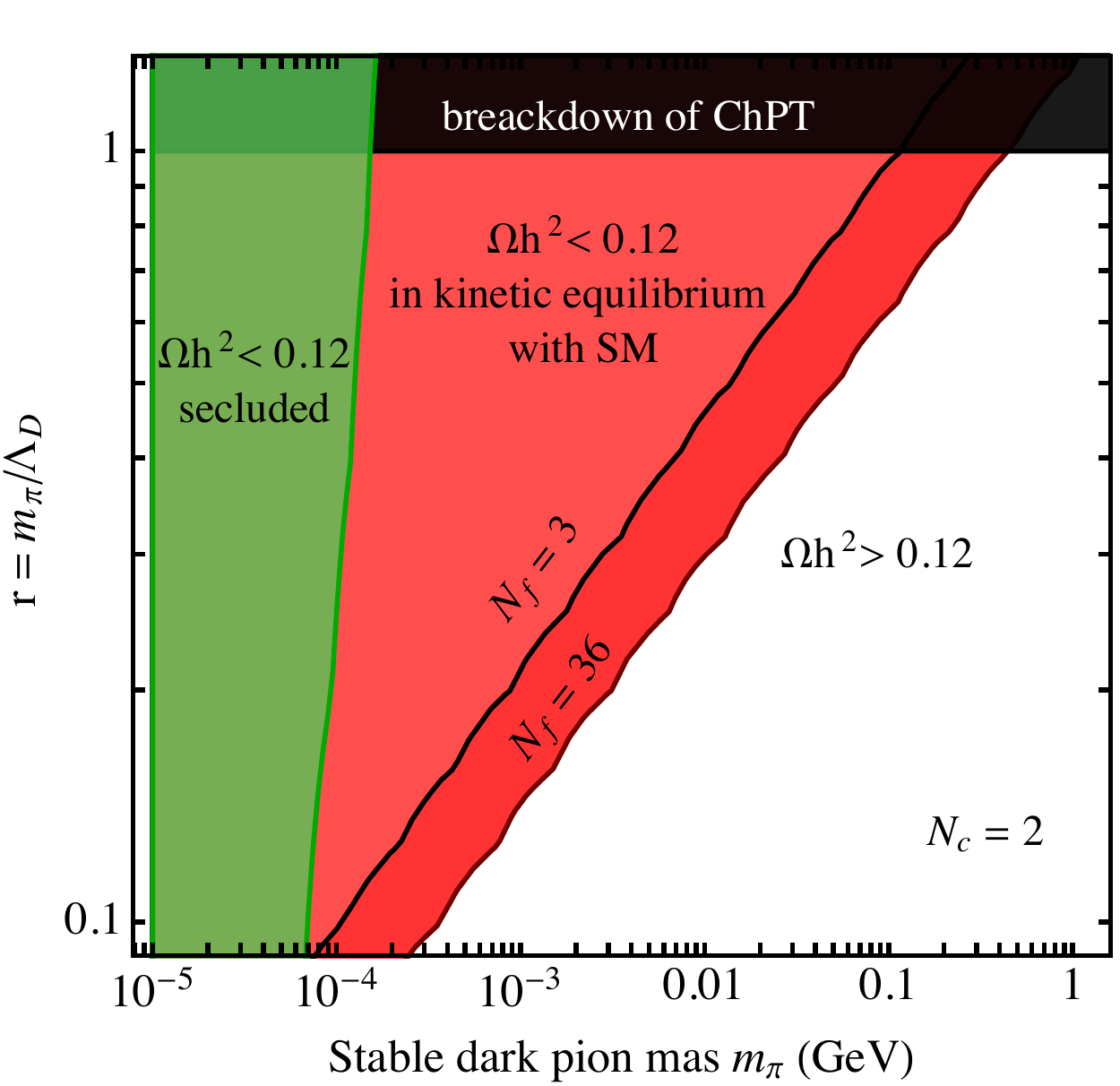}
\caption{\label{fig:RelicPionStable} The red shaded region describes the case that dark pions are in kinetic equilibrium with SM particles. It shows the parameter space within which the dark pion relic abundance is below the experimentally observed DM abundance. The region intersects with the black band, which indicates the breakdown of chiral perturbation theory at a pion mass order GeV. The green shaded region describes the scenario, where the dark pions are kinetically decoupled from the SM bath and show number-changing behavior and self-heating. Note that in this case the dependence on the pion to baryon mass ratio $r$ is much weaker. }
\end{center}
\end{figure}

Fig.~\ref{fig:RelicPionStable} shows the numerical result for the stable dark pion relic density for $N_{DC} = 2 $, $N_f =3$ and $N_f =36$, where $SU(N_f)_R \otimes SU(N_f)_R$ is broken to a diagonal $SU(N_f)_V$. The results superbly agree with the analytical estimate. 

There is an other important point which is crucial and very general to take into account here. The pion self scattering cross section is given by~\cite{Weinberg:1966kf}
\begin{align}
\sigma_{2 \rightarrow 2} = \frac{ 4 \pi^3 r^4}{m_\pi^2 } \frac{(N_f-1)(N_f+1) (3 N_f^4 - 2N_f^2 +6)}{N_f^2 (N_f^2-1)^2}\approx 12 \, \pi^3 \, r^4/m_\pi^2\,.
\end{align}
Note that this cross section only weakly depends on the number of fermion flavors. Observations of galaxy clusters provide upper bounds  on the DM self scattering cross section demanding that $\sigma_{2 \rightarrow 2}/m_\pi < \text{ cm}^2/\text{g} \approx 4.5 \,10^3 \text{ GeV}^{-3}$~\cite{1504.03388,1504.06576,1608.08630}.  Given the upper bound, we derived on the pion mass from the relic density constraint, we get a lower bound on this quantity $\sigma_{2 \rightarrow 2}/m_\pi  > 1.8 \times 10^4/(r^6 N_{DC}^2 N_f^3) \text{ GeV}^{-3}$, which implies that $r > 1.25\, N_{DC}^{-1/3}N_f^{-1/2}$.

This lower bound on $r$ indicates that at low values of $N_{DC}$ and $N_f$ the viable models are close to the regime where chiral perturbation theory breaks down. In fact~\cite{NNLO} demonstrated that at values of $r>0.3$ NLO and NNLO corrections increase the self scattering cross section beyond the experimentally viable values. A possible way to reconcile the stable pion DM scenario with data is to consider modes with large numbers of colors or fermion flavors $N_{DC} \geq 20$ or $N_f \geq 16$, such that $r<0.3$ and higher orders of the chiral expansion are sub-leading. It is intriguing that the allowed mass range for stable pions is in the sub-GeV regime, which is compatible with the pion lifetime estimates we performed in the introduction. Thus this scenario is self-consistent. 
 
\subsubsection*{Kinetically Decoupled Pions}

Even though Ref.~\cite{LaixScherrer} shows that self-heating secluded pions cannot be the DM of the universe, Ref.~\cite{Schmalz} discusses the possibility that a sub-dominant DM fraction of self-heating light particles is acceptable and might alleviate the observed tension in the matter power spectrum. To this end, the fraction has to be of order $1 \%$ of the total DM relic density.

 Fig.~\ref{fig:RelicPionStable} shows the numeric result for the secluded dark pion relic density as the green shaded region. Since in analogy to the baryon freezeout the Hubble rate has a complicated dark temperature dependence there is no such simple analytical expression for the final relic density. It is interesting to observe, that in this case the mass of the dark pions is expected to be between $10$ and $100$ keV, with a mild dependence on the ratio $r = m_\pi/\Lambda_D \approx m_\pi/m_\mathcal{B}$, with the dark baryon mass $m_\mathcal{B}$.  
 
 As shown above, in order to obtain a symmetric relic abundance of dark baryons in accordance with the observed DM relic density, those baryons have to have mass of order $100$ TeV, which implies an extremely small value of $r<10^{-10}$, which makes the pion self scattering cross section tiny. In the case that dark baryons are produced with an asymmetry and a smaller mass, the ratio $r$ can be much larger. However, since the dark pions are a subdominant DM fraction, the DM self-interaction constraint does not apply to them.


 \subsection{The Weakly Coupled Bound-state Limit } \label{sec:strongcoupling}

In this subsection, we consider the interaction regime, in which $M_Q \gg \Lambda_{D}$, thus bound states are perturbative objects similar to heavy quarkonia. In this regime, the annihilation takes place in two ways. 

\subsubsection{Dark Baryon Interactions}

\begin{enumerate}
\item Constituent annihilation which is dominant when the length scale corresponding to the baryon momenta is shorter than the naive size  of the colliding baryons $(m_\mathcal{B} v_{\rm rel} )^{-1}= p_\mathcal{B}^{-1} \ll R_\mathcal{B} \approx (\alpha_D C_N m_Q)^{-1}$, which implies that $v_{\rm rel} \gg \alpha_D C_N/N_{DC}$ with the group factor $C_N$ defined below. The cross section in this case is analogous to the free quark case since it is dominated by direct constituent annihilation and it is given by $\sigma_{\rm ann} v_{\rm rel} \approx  \kappa \pi \alpha_{\rm D}^2 /m_Q^2$ (with $\kappa$ a model dependent number of $\mathcal{O}(1)$).
\item Rearrangement annihilation, which is effective in the complementary regime i.e. $v_{\rm rel} \ll \alpha_D$. The process takes place when two baryons rearrange into a quarkonium state with $2 (N_{DC}-1)$ quarks emitting a meson $\bar{B} \,B \rightarrow \bar{Q}Q (N_{DC}-1) + \bar{Q}Q$. The $ (N_{DC}-1)  \bar{Q}Q$ state has baryon number zero and self-annihilates by stepwise meson emission. The rearrangement cross section is given by $\sigma_{\rm ann} \approx \pi R_\mathcal{B}^2/\sqrt{E_{\rm kin}/E_\mathcal{B}}$ and $\sigma_{\rm ann} v_{\rm rel} \approx \pi/(\sqrt{N_{DC}} C_N \alpha_D \, m_Q^2)$, where $C_N$ is the quadratic Casimir of the fundamental representation of the considered group \cite{BaryonDM}.  
\end{enumerate} 
 \begin{figure}[t]
\begin{center}
\includegraphics[width=0.45\textwidth]{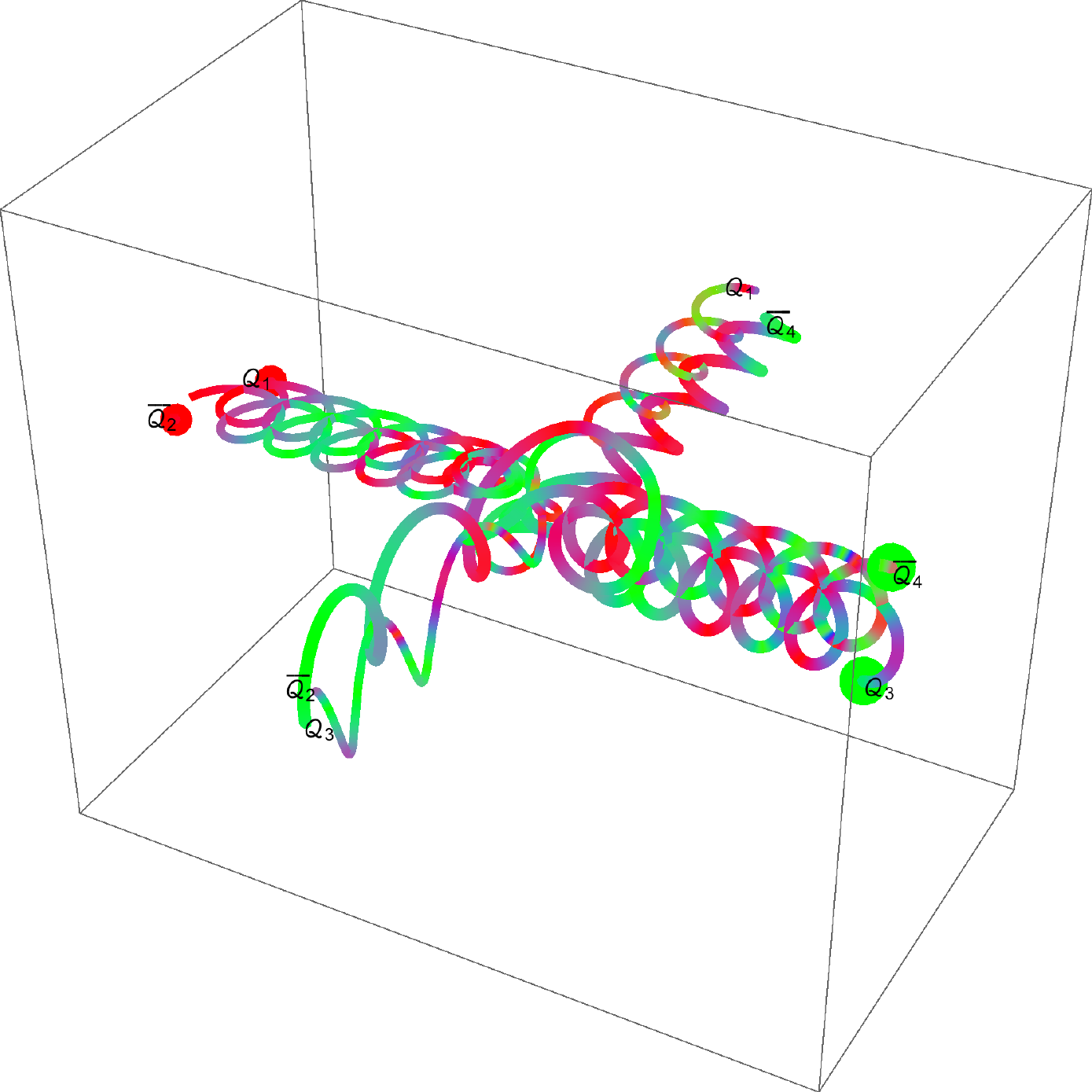}
\caption{\label{fig:ColorAnnihilation} Classical meson rearrangement of weakly coupled bound states. The quarks are in the fundamental representation of $SU(2)$.
The line color indicates the time-evolution of the bilinear color vector. The numerical study of this system confirms that the process is $s$-wave dominated and $ (\sigma_{\rm ann}\, v) \approx \text{ constant}$, given by Eq.~(\ref{eq:effectiverearrangment}).}
\end{center}
\end{figure}
The resulting cross section is thus a temperature dependent quantity, given by
\be\label{eq:effectiverearrangment}
 (\sigma_{\rm ann}\, v) \approx  \left\{
\begin{array}{ll}
 \frac{\kappa \pi \alpha_{\rm D}^2}{m_Q^2}  & \mathrm{;} \,  \qquad T_D > \frac{\alpha_D^2\, C_N^2 m_\mathcal{B}}{N_{DC}^2}\ ,\\
\frac{\pi}{\sqrt{N_{DC}} C_N \alpha_D \, m_Q^2}  & \mathrm{;} \, \qquad T_D < \frac{\alpha_D^2\, C_N^2 m_\mathcal{B}}{N_{DC}^2}\, .
\end{array}
\right.
 \ee
 The group factor can be computed in a given model, for example for an $SU(N_{DC})$ group it is $\kappa =( N_{DC}^4  -3 N_{DC}^3 +2 )/16\,N_{DC}^3$, assuming the quarks are SM singlets \cite{BaryonDM}.
 
The value of the rearrangement cross section quoted above can be estimated by numerical simulations of the classical rearrangement process, where the color charge of the dark quarks is taken into account. This is done by augmenting the Maxwell equations, by a classical evolution equation for the color vector $Q^a(t)$ with the adjoint index $a$, which is constructed from the bilinear $Q^a = c^\dagger_i \lambda^a_{ij} c_j$ with the color vectors of the fundamental fields $c_i$ and the corresponding group generators $ \lambda^a_{ij}$ ~\cite{Wong:1970fu}. The parallel transport equation dictates the evolution of this color vector $v \cdot D Q^a =0$~\cite{Drechsler:1981nc, 1611.03493}, where $v_\mu$ is the four velocity of the quark. This equation can be written and approximated in the non-relativistic limit as 
\be
\dot{Q^a_i}(t) = g_D f^{a b c} v^\mu A_\mu^b\, Q_i^c \approx g_D f^{a b c} A_0^b\, Q_i^c\,.
\ee
where $f^{abc}$ is the structure coefficient of the gauge group.  At leading order in the gauge coupling the equation for the potential in the Lorentz gauge is just $\Box A^\mu_a = g_D \, J^\mu_a$, where the sources are $J^\mu_a = \sum_i \int d\tau Q_a(\tau) v^\mu_i \delta(x - x_i(\tau))$. We can use the Lienart-Wiechert ansatz and get at leading non-relativistic order
\be
A_0^b = \frac{g_D}{4 \pi} \sum_{j\neq i}\, \frac{Q_j^b}{| \vec{r}_i - \vec{r}_j|}\,. 
\ee
With the Lorentz force equation $m_Q \ddot{\vec{x}}_i = g_D Q^a_i \, \vec{\partial} A_0^a$ for the positions of the color charges we obtain a closed system of differential equations. As an example we show a simulated classical meson, anti-meson rearrangement event in Fig.~\ref{fig:ColorAnnihilation}.

\subsubsection{Glueball Interactions}

The glueball spectrum is well known from lattice studies~\cite{GBspectrum} and contains the $0^{++}=S$ state as lightest particle. The interactions among the glueballs can be described by a low energy effective Lagrangian, obtained in the large $N_{DC}$ expansion \cite{Sigurdson, Cornwall:1983zb, Migdal:1982jp}
\begin{align} \label{eq:effLagGB}
\mathcal{L}_{\rm eff} \supset \frac{1}{2} (\partial S)^2 - \frac{a_2}{2!} \LDC^2 S^2 - \frac{a_3}{3!} \left(  \frac{4 \pi}{N_{DC}}\right) \LDC S^3 - \frac{a_4}{4!} \left(  \frac{4 \pi}{N_{DC}}\right)^2 S^4 + ...\,, 
\end{align}
with $a_i$ of order unity. From this Lagrangian the cross section for the $3 \rightarrow 2$ processes can be computed 
\begin{align}
\label{eq.32GB}
\sigma_{3 \rightarrow 2} v^2 \approx \frac{1}{(4 \pi)^3} \left(  \frac{4 \pi}{N_{DC}}\right)^6 \frac{1}{m_{\rm GB}^5}\,.
\end{align}

\subsubsection{Glueball Decay}

Glueballs are not protected from decay by any accidental symmetry. Therefore, generically they will have a limited lifetime. If the theory contains quarks, which carry electroweak quantum numbers, effective operators will be induced. 
If no Yukawa couplings with the standard model fermions are present, the leading operator will be of dimension eight \cite{BaryonDM}
\begin{align}
\mathcal{O}_8 = \alpha_{\text{EM}} \alpha_D G^a_{\mu \nu} G^{\mu \nu \, a} F_{\lambda \sigma} F^{\lambda \sigma } \,,   
\end{align}
with the coefficient induced by the quark loop
\begin{align}
c_8 = \frac{T_D \left( T_2 + d_2 Y \right) }{60 \, m_Q^4} \,.
\end{align}

For the simple model, we have introduced in the previous section, the only quarks in the theory are in the $Q = (3, 1, 3,0)$ of $SU(3)_D \times SU(3)_c \times SU(2)_L \times U(1)_Y$ and thus $T_D =1/2$, $T_2 = 2$, $d_2 =3$ and $Y=0$. 
The decay with of the glueballs is therefore 
\begin{align}
\label{eq:glueballdecay}
\Gamma_G =  \frac{\alpha_D^2 m_{\rm GB}^3 f_{G}^2}{3600 \pi \, m_Q^8} \left( \alpha_{ \text{EM} }^2 + 2 \alpha_2 \alpha_{ \text{EM} } \cos^2\left( \theta_W \right)  \sqrt{1 - \frac{ m_Z^2}{m_{\rm GB}^2}} \right.   \nonumber \\
\left.  +  \alpha_2^2 \cos^4\left( \theta_W \right)   \sqrt{1 - \frac{4 m_Z^2}{m_{\rm GB}^2}}  +  2\, \alpha_2^2   \sqrt{1 - \frac{4 m_W^2}{m_{\rm GB}^2}}    \right) \,.
\end{align}
The constant $f_G$ can be extracted from lattice data as in \cite{Junkewich} $f_G := \bra{0} \text{Tr } G_{\mu \nu } G^{\mu \nu } \ket{0^{++}}\approx  3 m_{\rm GB}^3/ 4 \pi \alpha_D$ . 
In the case of heavy glueballs with $m_{\rm GB} \gg m_W$, the kinematical bracket separating the $\gamma \gamma$, $Z \gamma$, $ZZ$ and $WW$ channels respectively, simplifies to $3 \, \alpha_2^2$.

\subsubsection{Phase Transitions}

 In this section, we will show how a first-order phase transition can lead to a dilution of frozen out relic particles~\cite{0909.1317}.  Numerical simulations of pure gluon dynamics indicate the presence of a first-order confinement phase transition for $N_{DC}$ larger than two. We can use this analogy for the dark scenario with heavy vector-like quarks. Here we are interested in the discontinuity of entropy density, which would affect the abundance of frozen out relics. 
 
 The entropy of the Yang-mills sector just above the critical confinement temperature is given by $s_D = 4 \pi^2/90 \, g_g T_D^3 - \partial B_0/ \partial T_D$, where $g_g$ is the number of gluons.   The non-perturbative contribution is encoded in the factor $B_0$. This factor can be modeled with the MIT bag model \cite{BagModel} and ascribes constant vacuum energy density to the interior of hadrons. However, just above the phase transition, the plasma can be considered as so dense, that $B_0$ is a universal global vacuum energy and thus does not affect the entropy function. The entropy density before the phase transition is thus $s_+ \approx s_g \approx  4 \pi^2/90 \, g_g T_D^3$.   
 
 After the confinement, the lightest degree of freedom is the $0^{++}$ glueball with a mass of $m_{\rm GB} \approx 7 \LDC $. Therefore, at $T_D< \LDC$ the glueballs are non-relativistic and have an exponentially suppressed contribution to the entropy density

 \begin{align}
   s_G  = \frac{ e^{-\frac{m_{\rm GB}}{T_D}} \sqrt{m_{\rm GB}^5 T_D}}{2 \sqrt{2} \pi^{3/2}} \approx s_-.
\end{align} 
 
If the dark sector dominates the energy budget of the universe at the phase transition i.e. if $\LDC < T_E$ a dilution effect will take place.  Since we argue that the non-perturbative entropy generating contributions can be neglected the change in entropy densities is compensated by an expansion of the universe i.e. $s_+ a_i^3 = s_- a_f^3$. The increase in the volume leads to a dilution of any relic already frozen out at the phase transition with the dilution factor given by    
 
  \begin{align}
D = \frac{s_{+}}{s_{-}} \approx \frac{4 \sqrt{2} \pi^{7/2}}{ 45} \frac{(N_{DC}^2 -1) T_{\rm conf.}^{5/2}  \, e^{m_{\rm GB}/T_{\rm conf.}}  }{m_{\rm GB}^{5/2} }  \,.
 \end{align}
 
 The confinement temperature can be approximated by $T_{\rm conf.} \approx \left(\frac{B_0 \, 90}{ \pi^2 \, (N_{DC}^2-1)  } \right)^{1/4}$, where $B_0 \approx \LDC^4$. For the case of $N_{DC} = 3$ the phase transition leads to a dilution factor $D \approx 400$.
 
 A further interesting aspect of this scenario is that during this first-order phase transition dark baryons will be accumulated up in clusters at the collapse points of the false vacuum. In case of an asymmetry in the dark sector, those DM clusters will persist until the present day.

 \begin{figure*}[t]
\centering
\subfloat[ \label{fig:WeakRelicD} No effect of the phase transition, since long lived or stable glueballs are assumed.]
{\includegraphics[width=0.45\textwidth]{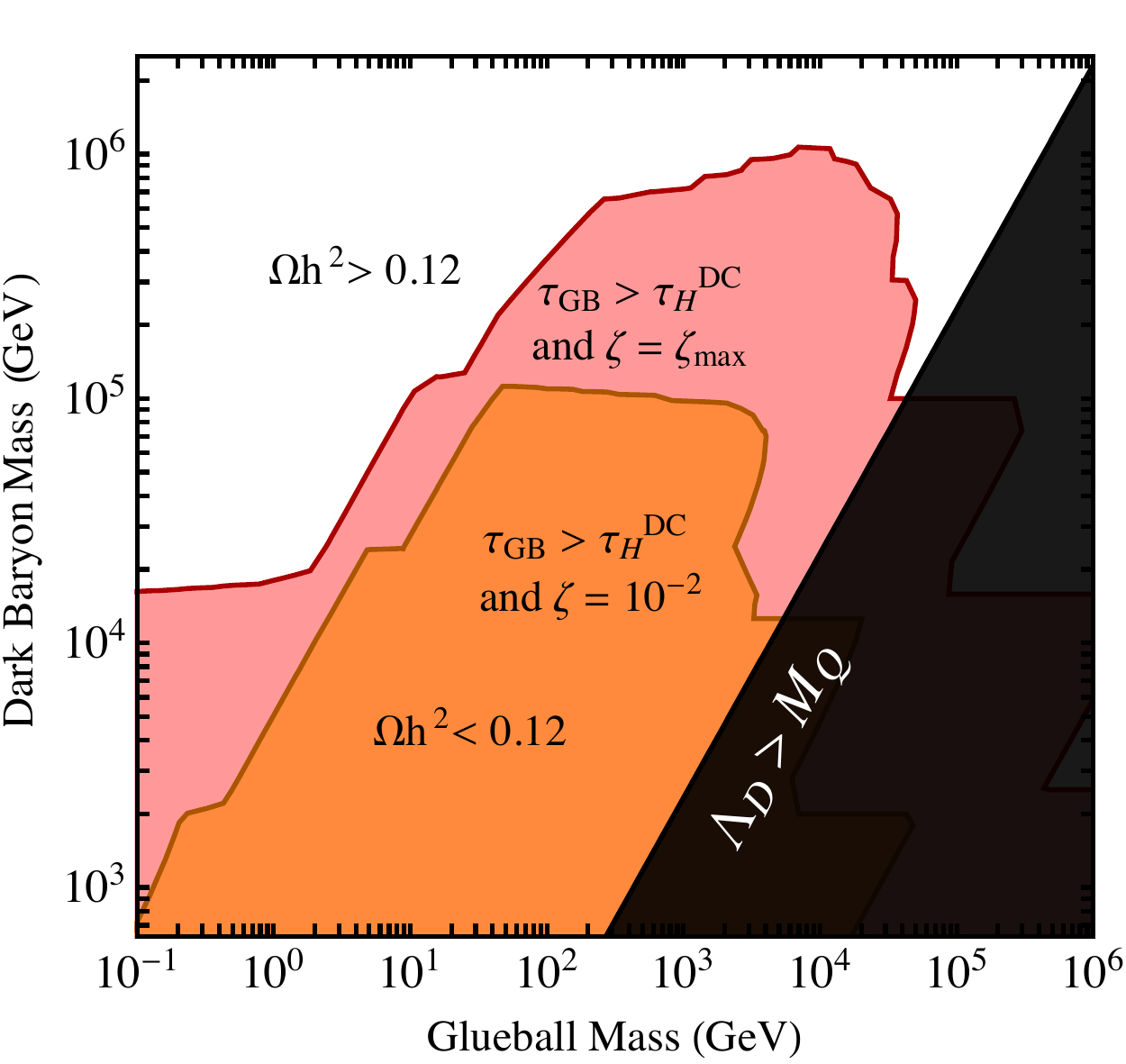}}
\hfill
\subfloat[ \label{fig:WeakRelicDwithD} Effective dark baryon dilution at the phase transition, which is first order in this case.]
{\includegraphics[width=0.45\textwidth]{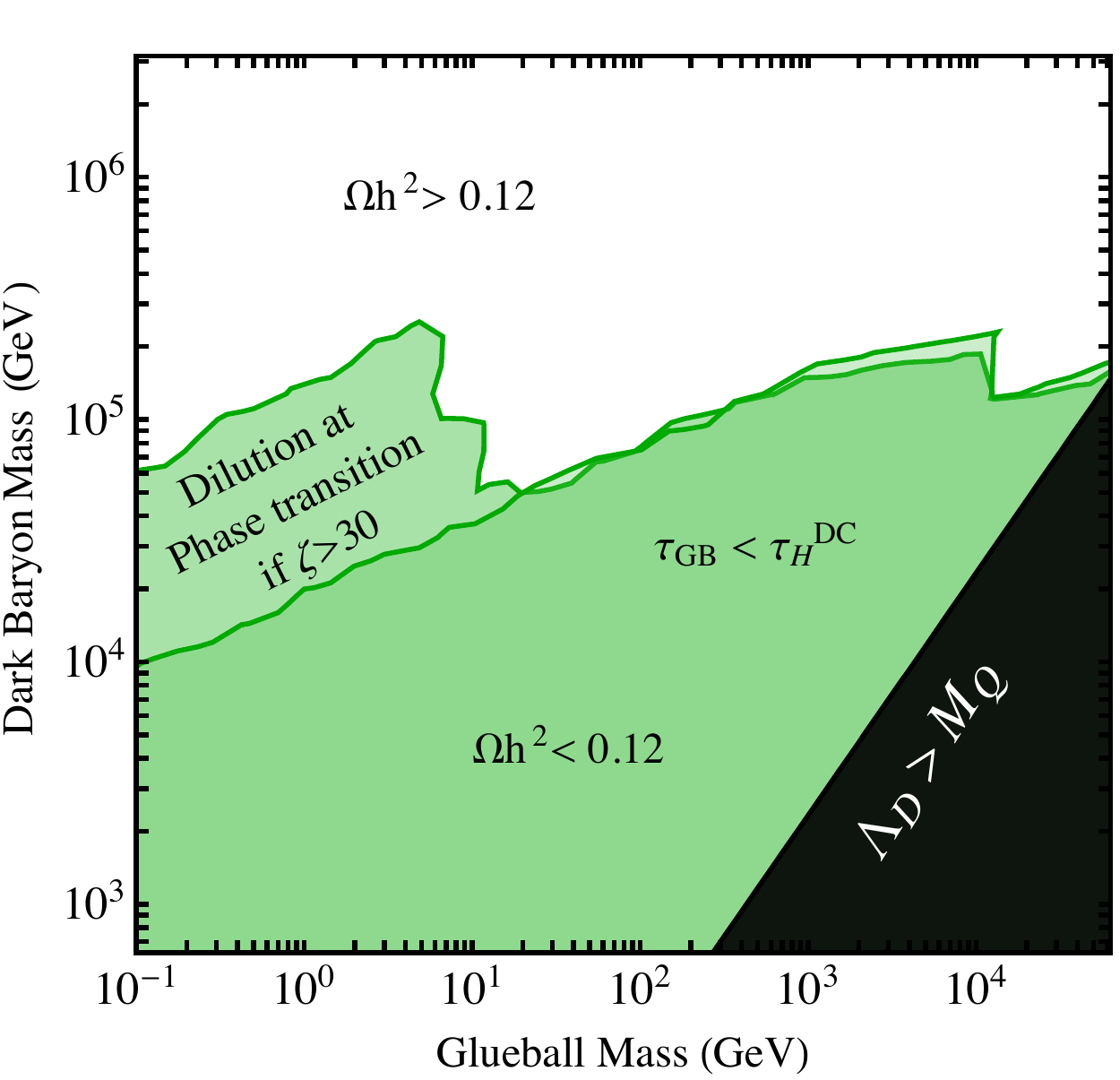}}
\caption{\label{fig:RelicDGBnodilution} The model parameter space allowed by the relic density constraint in the case of $\LDC \ll M_Q$. So far it is assumed that no entropy dilution is happening. The model considered has $N_{DC} = 3$. The left panel shows the relic abundance in the case that $\tau_{\rm GB} < \tau_\Lambda$ i.e. that glueballs decay immediately at formation. This leads to an effective thermalisation of the system and is the most predictive scenario. In the right panel the situation is considered, that glueballs are stable during the freeze-out process and the final relic abundance is sensitive to the entropy ratio at high temperatures $\zeta$. }
\end{figure*}
 \begin{figure}[t]
\begin{center}
\includegraphics[width=0.45\textwidth]{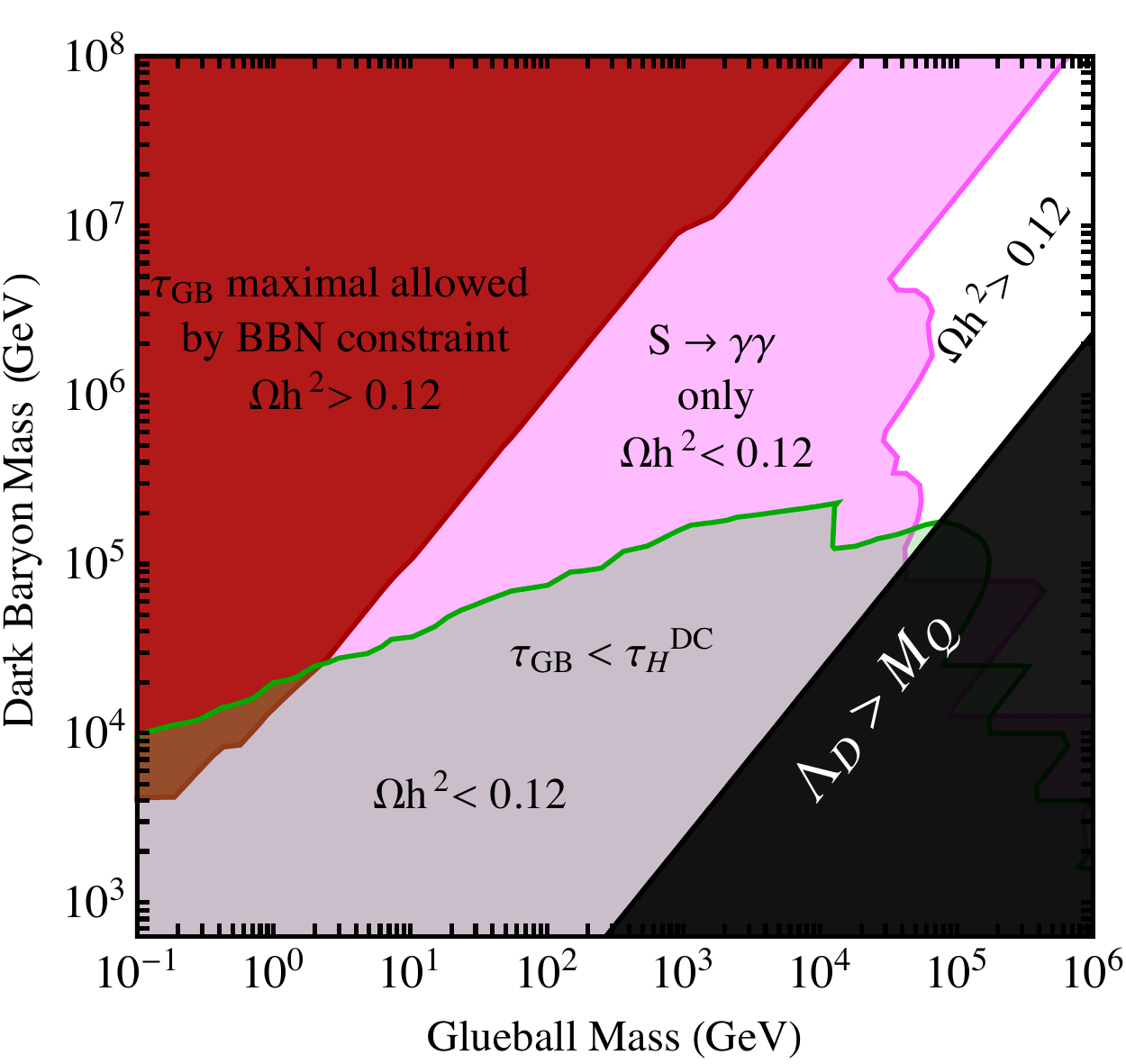}
\caption{\label{fig:RelicDGBdiluted} The model parameter space allowed by the relic density constraint in the case of $\LDC \ll M_Q$. The figure considers the entropy dilution, which results from the non-relativistic glueball decay. This dilution takes place if the glueballs dominate the energy budget of the universe. The contours show the extreme cases A) of glueball lifetime being maximal allowed by BBN i.e. $\tau_{\rm G} \approx 1 \text{s}$. And B) considers the minimal scenario, in which the only quark present in the model and forms baryons is in the $3$ representation of $SU(2)_L$, which leads to the glueball decay $S \rightarrow \gamma \gamma$, shown in Eq.~(\ref{eq:glueballdecay}).  The allowed parameter space for $\tau_{\rm GB} < \tau_\Lambda$ is shown for comparison. The model considered has $N_{DC} = 3$.}
\end{center}
\end{figure}

\subsubsection{Dark Baryon Abundance}

Equipped with our formalism the relic density of weakly coupled dark baryons can now be computed. We begin by exploring the parameter space keeping the SM quantum numbers general and varying the glueball lifetime as a free parameter and then focusing on the predictions of a concrete simple model. 

Fig.~\ref{fig:WeakRelicD} shows the relic abundance of dark baryons, in the case that the dark glueballs are stable at the baryon freezeout.  The final abundance depends on the initial entropy ratio $\zeta$, which is bounded from above by the number of SM degrees of freedom, assuming that the dark and SM sectors were in thermal contact at some higher temperature. 

Fig.~\ref{fig:WeakRelicDwithD} shows the relic abundance under the assumption that dark glueballs are not stable at the time of the dark baryon freezeout and establish a partial thermal link between the systems. Furthermore, we show that given an entropy ratio $\zeta > 30$ at the first-order phase transition in the case of low mass dark glueballs additional dilution takes place and leads to larger dark baryon masses. 

Fig.~\ref{fig:RelicDGBdiluted} shows the allowed parameter space in the concrete model with only one quark in the adjoint representation of the weak $SU(2)_L$ group. The glueballs decay according to Eq.~(\ref{eq:glueballdecay}). Given the model parameters, the glueballs decay after dark baryon freezeout and strong entropy dilution takes place. This is indicated by the purple contour which goes up to dark baryon masses, which are much larger than the expected unitarity bound~\cite{UniBound}. The maximal entropy diluted relic density is shown by the red contour. It takes place if the glueball lifetime is assumed to be $\tau_{\rm GB} \approx 1 \text{ s}$, so as large as it can get, given the BBN constraint. 
For comparison, the dark baryon relic abundance result with short-lived glueballs is shown in green. Note that the provided upper bounds on the DM mass from the requirement $\Omega h^2 <0.12$ also apply if a DM asymmetry is present. Since in that case, the symmetric component has to annihilate away to at least that value in order not to overclose the universe.

\subsubsection{Stable Glueball Abundance}

When dark glueballs are long lived on cosmic time scales, they can be DM candidates as well. The effective Lagrangian for the glueball interactions is not based on approximate global symmetry but it is a result of a $1/N_{DC}$ expansion and works best at large number of colors~\cite{Sigurdson}. The glueball self-scattering cross section can be computed from Eq.~(\ref{eq:effLagGB}) and is directly related to the $3 \rightarrow 2$ reaction cross section
\begin{align}
\label{eq:22GB}
\sigma_{2 \rightarrow 2} \approx \frac{1}{16 \pi } \left( \frac{4 \pi}{N_{DC}} \right)^4 \frac{1}{m_{\rm GB}^2} \approx \left( \frac{N_{DC}}{8 \pi} \right)^2 m_{\rm GB}^3\,   \sigma_{3 \rightarrow 2} v^2 \,.  
\end{align} 
The associated relic abundance can be computed in two distinct regimes.

\subsubsection*{Dark Glueballs in Kinetic Equilibrium}

The same logic concerning kinetic equilibrium used earlier for dark pions applies also to dark glueballs, thus the stable glueball is only a good DM candidate if kinetic equilibrium is maintained~\cite{1602.00714}. 

The structure of the Boltzmann equation is given by Eq.~(\ref{eq:32boltzmanngeneral}) and thus identical to the dark pion freeze-out case. The only difference is that the $3 \rightarrow 2$ annihilation cross section for glueballs is temperature independent. Therefore, the function in Eq.~(\ref{eq:32boltzmanngeneral}) is $f(z) = z^{-5}$ and hence the asymptotic number density is given by ($n=5$): 
\begin{align}
Y_{\rm GB}^\infty = \frac{n_{\rm GB}}{s_{\rm SM}} \approx \frac{2}{\sqrt{\lambda}} \left(  \frac{1}{(z^{\rm GB}_c)^n}+ \frac{8}{n-1} \frac{1}{(z^{\rm GB}_c)^{n-1}} \right)^{-1/2} \approx \sqrt{\frac{2}{\lambda}} (z^{\rm GB}_c)^2\, \text{  and  }\Omega_{\rm GB} \approx \frac{s_{\rm SM}^0 m_{\rm GB} Y_{\rm GB}^\infty }{\rho_c}\,.
\end{align}

Given the strict relations between the glueball mass and the number-changing $3 \rightarrow 2$ cross section Eq.~(\ref{eq.32GB}), we have a concrete prediction $\Omega_{\rm GB} \approx 8.1 \, (m_{\rm GB}/\text{GeV})^{3/2} N_{DC}^{3}$. This implies an upper bound on the glueball mass $m_{\rm GB} < 0.1/ N_{DC}^{2} \text{ GeV}$ translating into $\sigma_{2 \rightarrow 2}  > 5 \, \times 10^4 \text{ GeV}^{-2}$ or equivalently $\sigma_{2 \rightarrow 2}/m_{\rm GB} > 5 \, \times 10^5 N_{DC}^2\, \text{ GeV}^{-3}$. This self scattering cross section violates the observational bound of $\sigma_{2 \rightarrow 2}/m_{\rm GB}  < 4.5 \,\times 10^3 \, \text{ GeV}^{-3}$ by more than two orders of magnitude. Therefore, stable dark glueballs in kinetic equilibrium with the SM plasma can not be the dominant DM component. 

\subsubsection*{Decoupled Dark Glueballs and Multicomponent Dark Matter}

The other DM avenue for dark glueballs is that they are not in kinetic equilibrium with the SM plasma and have their temperature evolution, which leads to a prolonged phase with non-vanishing pressure. The only phenomenologically viable scenario, in this case, is a multi-component dark sector~\cite{arXiv:1207.3318, 1007.4839}. The glueballs comprise in this case only a subdominant DM fraction with a much larger free-streaming length than the dominating DM candidates. This phenomenon could resolve some large scale structure problems if the glueballs make up for about $1\%$ of the total DM relic abundance~\cite{Schmalz}. 

\begin{figure}[t]
\begin{center}
\includegraphics[width=0.45\textwidth]{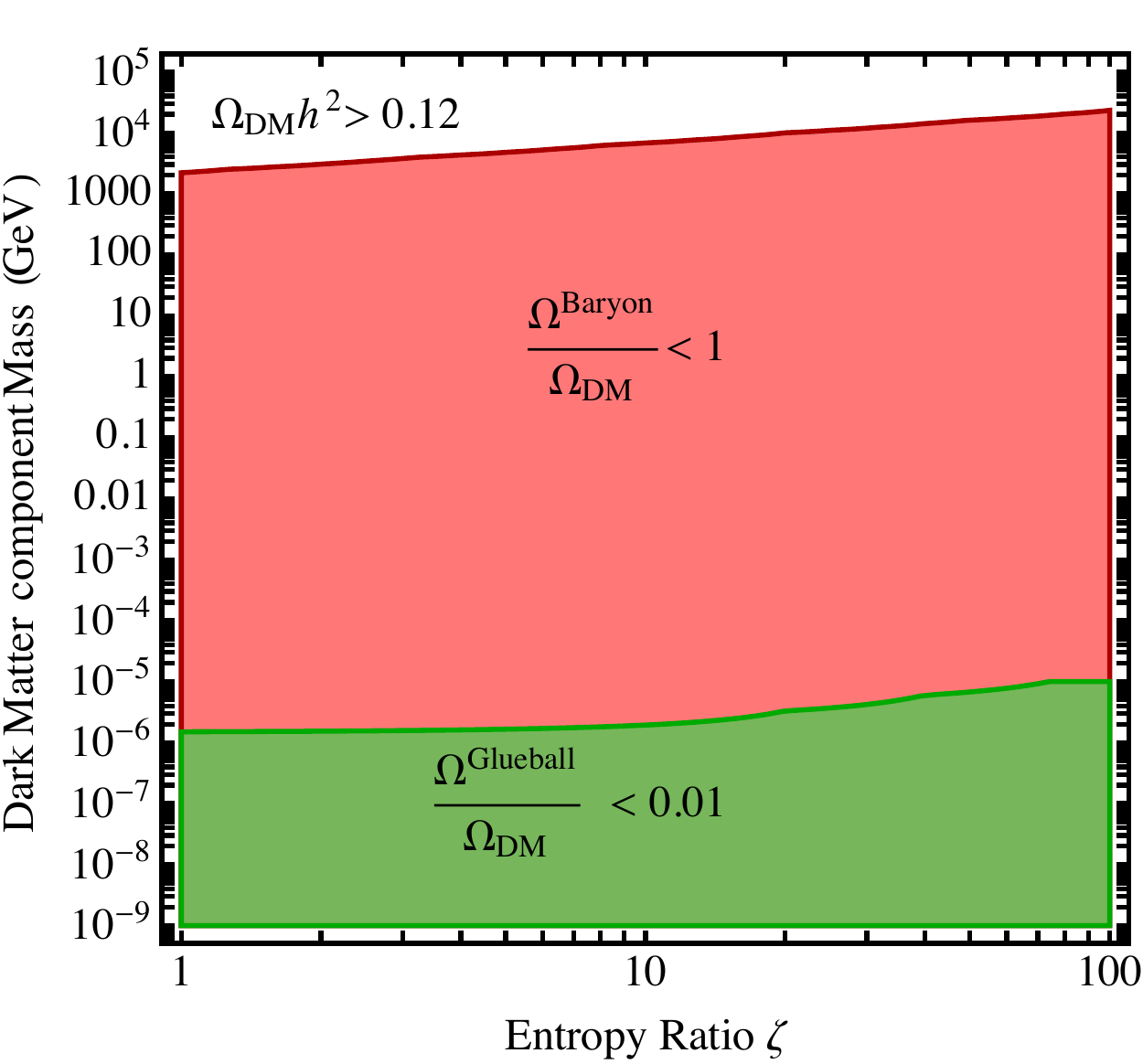}
\caption{\label{fig:MulticomponentDM} The model parameter space allowed by the relic density and structure formation constraint in the case of $\LDC \ll M_Q$ and long-lived glueballs. The resulting model features are two-component DM system with TeV scale dark baryons, which make up about $99\%$ of the DM and a subdominant keV scale glueball population (about $1\%$ DM abundance). This self-heating glueball component can help to resolve observational tension in the matter power spectrum~\cite{Schmalz}. The model considered has $N_{DC} = 3$.}
\end{center}
\end{figure}

The explicit numerical solution of the Boltzmann equations leads to a dark glueball mass between $100$ keV and $10$ MeV, depending on the initial entropy ratio $\zeta$. A few comments are in order 
\begin{enumerate}
\item The low mass glueballs are expected to have sizeable self scattering cross sections. However, the cluster constraints do not exclude them, since they are a subdominant DM component. 
\item Such low masses indicate a dark confinement scale close to or even below $\Lambda_{\rm QCD}$. 
\item In contrast to previous scenarios, the low confinement scale is phenomenologically acceptable, since the dark glueballs do not decay and do not affect BBN in this scenario.
\item When dark glueballs constitute a $1\%$ fraction of the DM abundance motivates a low confinement scale and simultaneously predicts a mass range of a few TeV for the dark baryons emerging as bound states of heavy dark quarks, see Fig.~\ref{fig:MulticomponentDM}.
\end{enumerate}

This scenario is special from an experimental point of view since glueballs are produced during annihilation of dark baryons. Since the glueballs are long-lived and feebly interacting with SM matter those decay channels are experimentally invisible. Therefore, the detectability in this scenario depends strongly on the specific SM quantum numbers of the dark quarks that will lead to subdominant annihilations of dark baryons into SM final states. Another possibility is that the dark glueballs, despite being long-lived, slowly decay leading to detectable signals in indirect detection experiments.  We will now come to the discussion of experimental signatures of composite DM models. 


\section{Experimental Constraints} 
\label{sec:constraints}

The composite DM models we discussed here can be very broadly divided into two classes:
\begin{itemize}
\item Type I: Theories with composite DM particles that carry residual SM interactions. This is the case, for example, of one weakly charged dark quark in which the lightest baryon is by construction a multiplet of the weak force~\cite{AccidentalDM} but still neutral under QCD. Another possibility are models featuring dark quarks charged under QCD, which lead to residual induced chromo-electric interactions~\cite{ColoredDM, ColoredRelics}. Another time-honored example is a dark technibaryon \cite{0809.0713} which has still residual SM interactions due to its SM charged constituents. 
\item Type II: Theories with composite DM particles,  are total SM singlets. This can be realized in a model where the lightest dark quarks are SM singlets and heavier ones carry SM quantum numbers leading to LCP decays.
\end{itemize}

Here we focus on the indirect detection experiments since the dark-baryon-matter interactions, important for direct searches, are only present in models of type I and are highly model dependent \cite{BaryonDM}. In general, viable SM quantum numbers for dark baryon systems were classified in~\cite{AccidentalDM}.

In both type I and II models the LCPs are unstable and therefore we expect annihilation processes of  $B \bar{B} \rightarrow  \text{ LCPs} \rightarrow$ {several} { SM-particles}.  The cascade depth and the type of final state SM particles are model dependent. However, since photons will be present for any SM final state experiments such as FermiLAT~\cite{FermiLAT} and H.E.S.S.~\cite{1711.08634}, as well as Planck Satellite~\cite{1507.02704} observations of the CMB~\cite{1506.03811} play an important role to discover these models \cite{0811.4153}. Shortly, the upcoming CTA experiment will probe the target cross section regions very efficiently already with the data of a single dwarf galaxy~\cite{1508.06128}. Furthermore, the details of the spectral shape are less relevant for an order of magnitude limit, because the spectrum is sufficiently softened by cascade decays \cite{StatyerCascade}. 

 \begin{figure}[t]
\begin{center}
\includegraphics[width=0.45\textwidth]{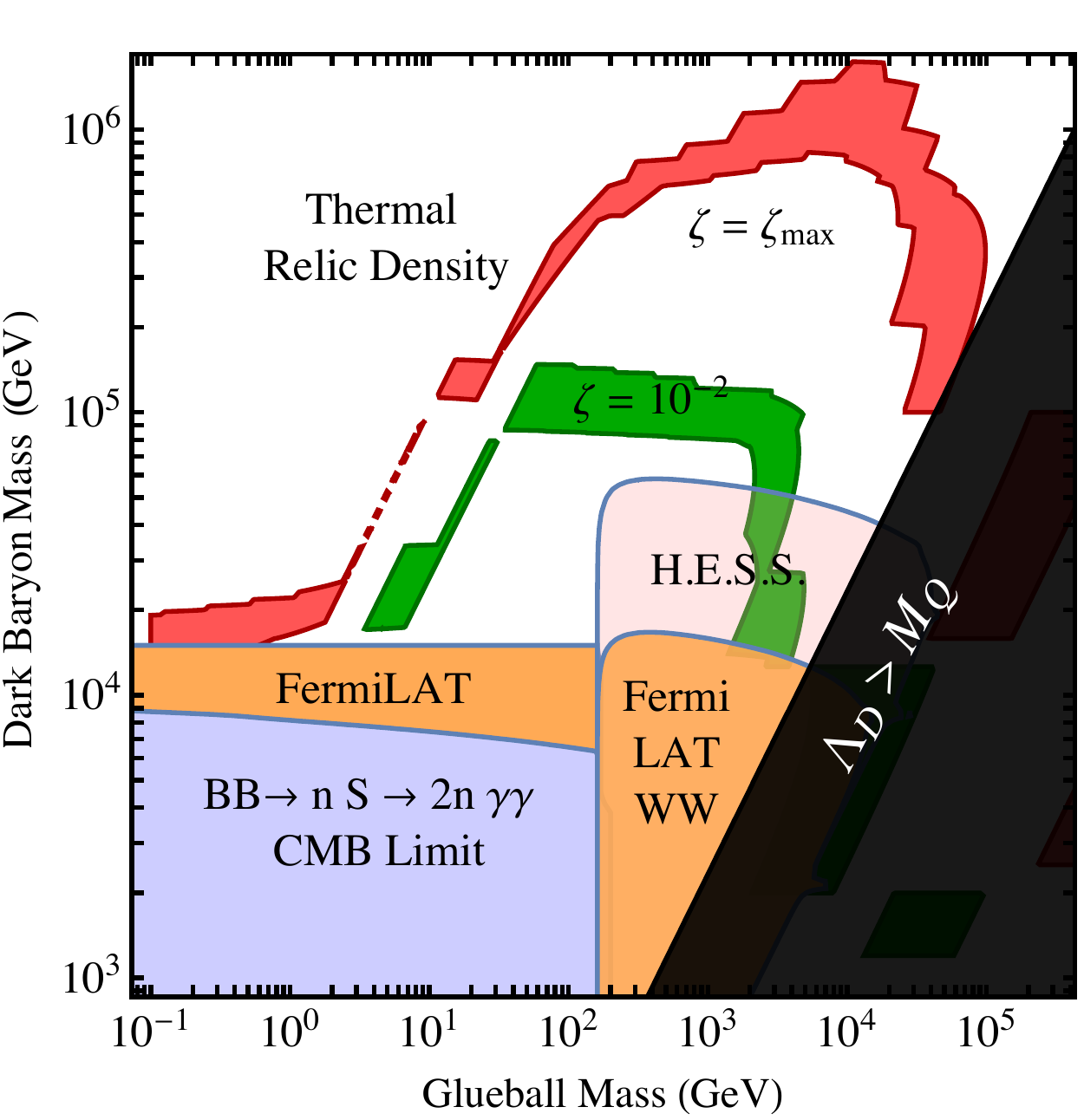}
\includegraphics[width=0.45\textwidth]{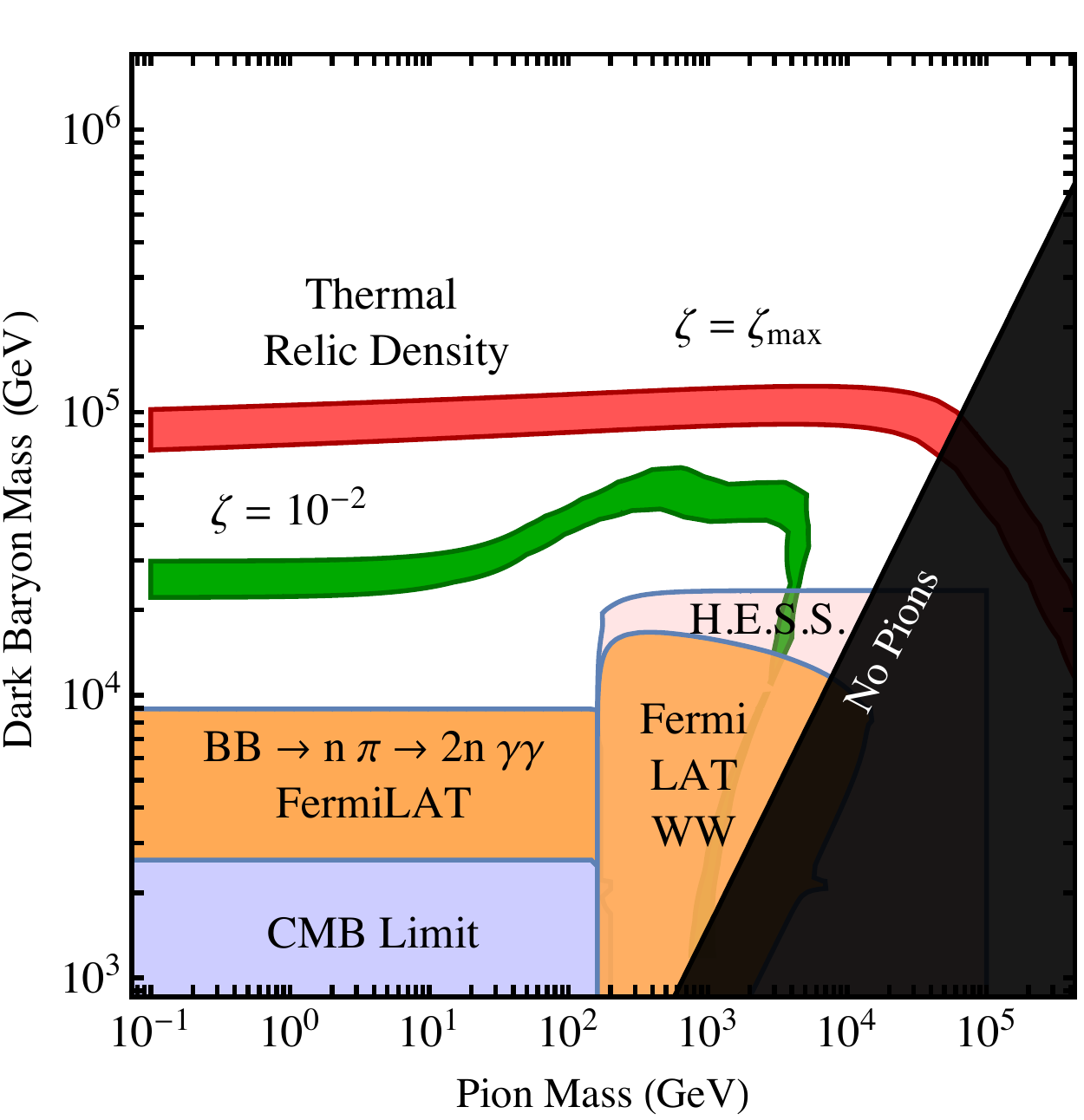}
\caption{\label{fig:limits}  The experimental limits on a dark-gauge sector with minimal charge assignment. The dark quarks are in the adjoint representation of $SU(2)_L$. In the left panel, we show the expected relic abundance for the weakly coupled baryons   \cite{BaryonDM}. In the right panel, we show the parameters expected for thermal production when the dark baryons are strongly coupled bound states. The most stringent limits for light LCPs are inferred from the FermiLAT experiment and Planck observations of the CMB \cite{SlatyerCMB}. Those limits are particularly insensitive to the cascade details and DM distribution, as only the total energy deposit matters. For heavier LCPs the H.E.S.S. limits are dominant, however, those limits are obtained under the assumption of an Einasto DM profile and would become weaker with a strongly cored DM profile. }
\end{center}
\end{figure}
\begin{figure*}[t]
\centering
\includegraphics[width=0.45\textwidth]{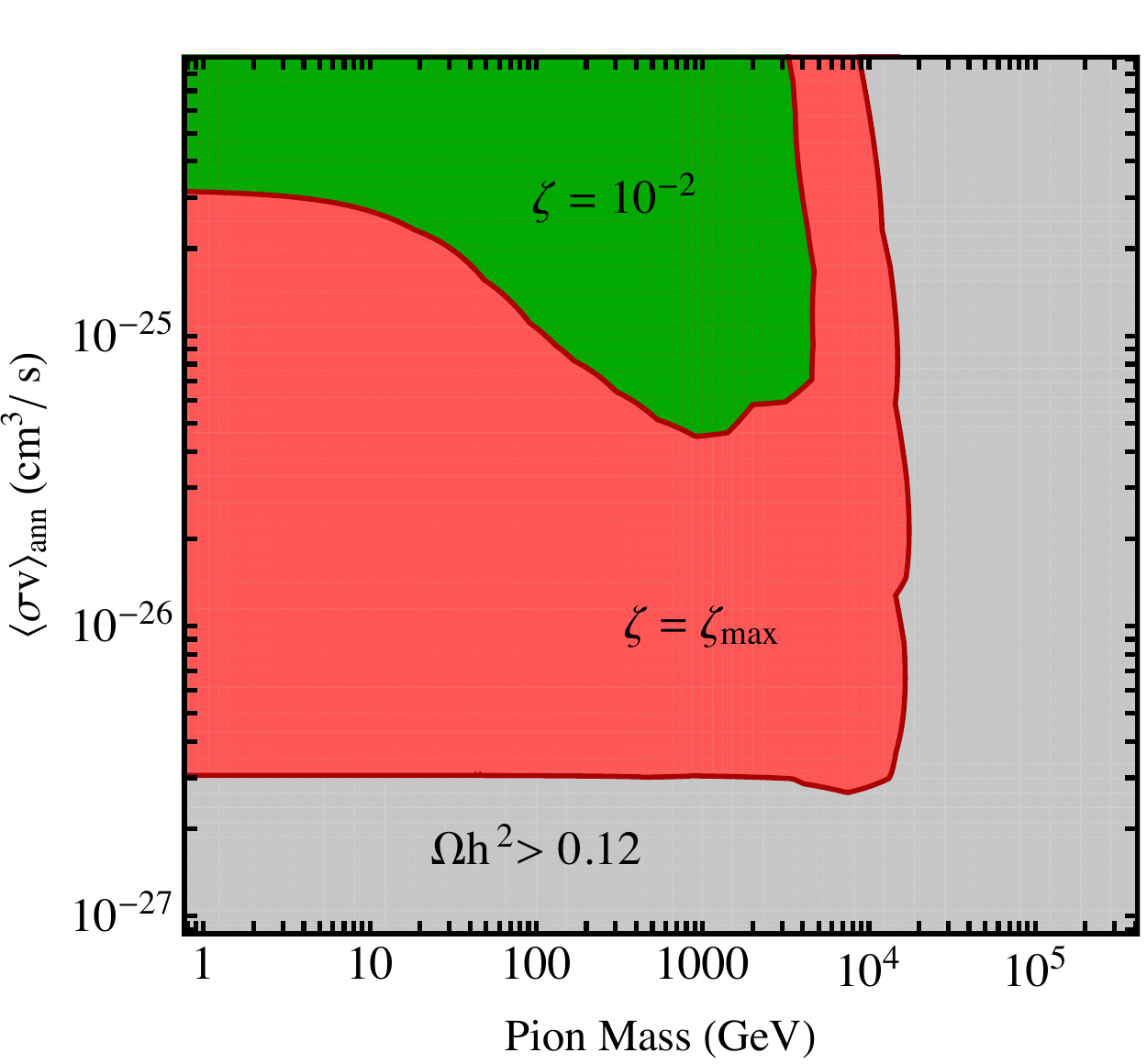}
\caption{\label{fig:target} The target cross section expected from the thermal freezeout in the strongly coupled baryon model as a function of the pion mass. While in the low pion mass regime the standard value of the annihilation cross section is recovered, larger pion masses lead to an enhanced annihilation cross section, increasing the detectability of this scenario.}
\end{figure*}

In Fig. \ref{fig:limits}, we present the limits for the minimal SM charge assignment of the dark quarks. Here, the spectrum of dark quarks contains a multiplet which transforms according to the adjoint representation of $SU(2)_L$. As discussed above this allows the dark glueballs or dark pions to decay to photons or heavy $SU(2)_L$ gauge bosons, if kinematically allowed.  Note that the limits on the dark baryon annihilation cross section will be in the same ballpark, even if the DM candidate baryon itself is a SM singlet given that the LCPs decay into photons.

In Fig.~\ref{fig:target} we show the expected annihilation cross section in the strongly coupled baryon model with SM singlet pions. This way of presenting the data is convenient to investigate the experimental reach into the model parameter space. We point out that at larger pion masses the expected cross section is significantly larger than the typically expected benchmark target value of $\langle \sigma v \rangle  \approx 10^{-26} \text{cm}^3/\text{s} $. This makes this region of parameter space particularly interesting from the point of view of detectability. Furthermore, we observe that decreasing the entropy ration $\zeta$ results in larger late time annihilation cross sections, which means the cross section parameter space is bounded from below and can be tested entirely given the necessary experimental sensitivity. 

Besides the pure dark sector effects in models of type I we also have residual interactions with SM particles that will generically lead to additional non-perturbative effects in the annihilation processes such as the Sommerfeld enhancement~\cite{Nima, Cassel, Hisano:2003ec, Hisano:2004ds, Hisano:2006nn} and bound-state formation~\cite{Petraki, Wise, PetrakiNew, Slatyer, CosmoDM, Braaten1, Braaten2, Braaten3, 0812.0559}.

The above scenario does not apply when considering long-lived glueballs yielding invisible dark baryon annihilation processes. Depending on the lifetime one could hope to experimentally test this scenario via the dark glueball decays into SM particles. Furthermore, since dark baryons keep annihilating into dark glueballs, which have considerable self-interactions, the dark matter self-interactions evolve with time. This could be an intriguing feature for late time dynamics of galaxies and galaxy clusters. We postpone a detailed study of this scenario to a later endeavor.  


\section{Conclusions}
 \label{sec:conclusions}

We investigated the thermodynamic history of confining dark sectors that can feature global accidental symmetries, leading to long-lived particles such as the proton for the Standard Model. 

Dark sectors can generically decouple from kinetic equilibrium with the Standard Model plasma and undergo a separate thermal evolution. Since the lightest composite particles being pions or glueballs unavoidably feature number-changing interactions, the secluded sector plasma heats up until the number-changing processes decouple. This behavior leads to a nonstandard dependence of the Hubble rate and the entropy evolution on the dark sector temperature. Thus to find the relic abundance, we have set up new Boltzmann equations, taking this atypical thermodynamical behavior into account.

We adopted our formalism to investigate two regimes of confining dark sectors, the strongly coupled bound-state regime, and the weakly coupled bound-state regime. We find several viable Dark Matter scenarios:
\begin{itemize}
\item Stable dark baryons, strongly coupled with masses around the 100 TeV scale.
\item Stable dark baryons, weakly coupled with masses between a few and 100 TeV.
\item Stable dark pions around the GeV scale and considerable self scattering cross sections.
\item Multicomponent Dark Matter scenarios with dark baryons at the TeV scale and a subdominant self-interacting glueball Dark Matter component at the $1\%$ level of the relic density with keV scale masses.
\end{itemize} 
Interestingly models featuring long-lived or stable light-dark particles can address issues related to the local dark-matter halos. Dark light glueballs and pions have considerably large self-interactions. Furthermore, in the case of a kinetically decoupled subdominant dark pion or glueball population, the self-heating due to number-changing processes in the dark plasma can resolve some experimental tension in the matter power spectrum.

Finally, we discussed how to experimentally test these models of composite Dark Matter. We argued that the most promising search strategy is indirect detection via space or ground-borne experiments. The crucial point is that these experiments are sensitive to the total annihilation cross section of Dark Matter particles into Standard Model final states via intermediate LCPs. We derive stringent limits from CMB observables, based on the total power injection. Furthermore, we show the sensitivity of experiments aimed at the Galactic Center. Remarkably, current indirect detection experiments can place bounds on the considered models, given the fact that the typical mass range of the dark baryons is mostly in the multi TeV regime. This lucky coincidence is related to the fact, that at the large dark baryon mass, effects of the dark plasma enhance the dark baryon production efficiency, which results in larger target cross sections. Thus we can hope that upcoming experiments, such as CTA, will be able to deeply cut into the composite dark matter parameter space. 


\section*{acknowledgements}
JS would like to thank Christopher Cappiello, Michael Duerr, John March-Russell and especially Michele Redi for helpful discussions.
 This work was supported by the CP$^3$-Origins centre, which is partially funded by the Danish National Research Foundation, grant number DNRF90.


\appendix

\appendix

\section{Summary of Dark Thermodynamics}
\label{appendixA}

\subsection{The Background Evolution}

We present the main formulas concerning the evolution of the Friedman-Robertson-Walker (FRW) universe, specified by the metric

\begin{equation}
d s^2 = - d t^2 + a(t)\left[ \frac{d r^2}{1-k r^2} + r^2 d \Omega^2 \right] ~.
\end{equation}

The source of spacetime expansion (or contraction) described by the scale factor $a(t)$ is taken to be a perfect fluid. This is described by particle current $N^{\mu}$ and stress-energy tensor $T^{\mu\nu}$ satisfying continuity equations:
\begin{equation}
\begin{split}
N^{\mu} &= n U^{\mu}, \quad \nabla_{\mu}N^{\mu} = 0 \implies \frac{1}{a^3} \frac{d}{dt} (n\,a^3) = 0 \quad \text{or} \quad \dot{n} + 3H\,n =0 ~,\\
T^{\mu\nu} &= (\rho+P)U^{\mu}U^{\nu} - P g^{\mu\nu}, \quad\nabla_{\mu} T^{\mu}_{\nu} = 0 \implies \dot{\rho} + 3H(\rho+P) = 0 ~.
\end{split}
\end{equation}

We denoted by $n(t)$ the particle number density and by $U^{\mu}=(1,0,0,0)$ the fluid four-velocity in the comoving frame. $H(t) = \dot{a}/a$ is the Hubble parameter. In general a relation between pressure $P$ and energy density $\rho$ of the fluid is required, for example:
\begin{equation}
P = w \rho ~ \begin{cases}
w = 0 \quad \text{non-relativistic matter}\\
w = \frac{1}{3} \quad \text{radiation}\\
w = -1 \quad \text{vacuum energy}
\end{cases}
\end{equation}
Pluggin the FRW ansatz in the Einstein equations one finds:
\begin{equation}
\begin{split}
H^2 &= \frac{8\pi}{3M_p^2} \rho - \frac{k}{a^2} \\
\dot{H} &= -H - \frac{4\pi}{3M_p^2} (\rho + 3P)\,.
\end{split}
\end{equation}
We will consider exclusively the case in which $k=0$. These equations uniquely determine the universe scale factor $a(t)$ with his own matter content.

\subsection{General Multi-fluid Thermodynamics}

In this section, we provide a compendium of formulas for equilibrium multi-fluid systems.
Considering first a single component fluid of particles, this is in \textit{kinetic equilibrium} if kinetic energy exchange through particle collision is efficient. In this case, the phase-space distribution is a Bose-Einstein (BE) or Fermi-Dirac (FD):
\begin{equation}
f(p) = \frac{1}{e^{\beta (E - \mu)} \pm 1} ~\sim~ e^{-\beta (E-\mu)} \quad\text{when}\quad \beta (E-\mu) ~\gg~ 1 ~.
\end{equation}

Relativistic particles have the dispersion relation $E = \sqrt{p^2 + m^2}$ and $\beta = 1/T$. The pressure, energy and particle density are obtained by
\begin{equation}
n = g \int \frac{d^3 p}{(2\pi)^3} f(p) ~, \quad \rho = g \int \frac{d^3 p}{(2\pi)^3} E(p) f(p) ~, \quad P = g \int \frac{d^3 p}{(2\pi)^3} \frac{p^2}{3E(p)} f(p) ~,
\end{equation}
with $g$ being the number of discrete degrees of freedom (spin, flavour, ecc..).
The entropy density of the system is obtained from
\begin{equation}
dS = \beta( dU + P dV - \mu dN)  \implies s = \beta(\rho + P - \mu n) ~.
\end{equation}
Useful expressions are obtained in the relativistic $\beta m \sim \beta \mu \ll 1$ limit :
\begin{equation}
n = \frac{\zeta(3)}{\pi^2} g T^3 \begin{cases} 1 & \text{BE} \\ \frac{3}{4} & \text{FD} \end{cases} ~, \quad \rho = \frac{\pi^2}{30} g T^4 \begin{cases} 1 & \text{BE} \\ \frac{7}{8} & \text{FD} \end{cases} ~, \quad P = \frac{1}{3} \rho ~, \quad s =\frac{4}{3T} \rho ~.
\end{equation}
For simplicity we will include the BE/FD factors in the definition of $g$.

In the non-relativistic regime $\beta m \sim \beta \mu>> 1$ we have instead:
\begin{equation}
n = g \left( \frac{mT}{2\pi} \right)^{3/2} e^{-\beta(m-\mu)} ~, \quad \rho = mn + \frac{3}{2} nT \sim mn \quad P = n T \sim 0 ~, \quad s = (m-\mu)\frac{n}{T} ~.
\end{equation}
Although we include the chemical potential in our treatment due to the presence of number-changing interaction, we will always assume to be far from the degenerate phase. In general as a consequence of continuity equations we have $\mu = \mu(T)$.

Suppose now that the ensemble contains many species of particles, and inter-species interactions are absent. Then every species thermalize separately, each one with its temperature and chemical potential $( T_i, \mu_i)$. 

In our treatment, it's convenient to work with dimension-less quantities, defined with respect to a given scale $M$.
In terms of $z=M/T,z_{\mu_i} = \mu_i/T$. The entropy density in relativistic and non-relativistic regimes are, respectively: 

\begin{equation}
s_i = \left(\frac{2\pi^2}{45}\right) g_i M^3 z^{-3} ~, \quad  s_i = \frac{g_i M^3 }{\sqrt{8\pi^3}}~ r_i^{3/2} \left(r_i z - z_{\mu_i}\right) z^{-3/2} e^{-(r_iz-z_{\mu_i})} ~.
\end{equation}

Where we introduced the mass ratios $r_i = m_i/M$.
If inter-species interactions are switched on, these can provide extra channels for thermalization (elastic processes) or species-changing interactions (inelastic processes). The balancing of the latter reactions for the time-reversed processes is governed by the chemical potentials. When the sum of chemical potentials of reacting particles are equal on both sides of the reaction the system reaches \textit{chemical equilibrium}, and the fraction of each species is then determined and not evolving in time.

A multi-fluid system that is both in kinetic and chemical equilibrium is said to be in \textit{thermal equilibrium}. In this case, every species follows the equilibrium distribution with a common temperature of $T_i = T$.

When all the species are in thermal equilibrium the total entropy density is 

\begin{align}
s_{tot} &= \left(\frac{2\pi^2}{45}\right) M^3 z^{-3} \left[ \sum_{r_i z << 1} g_i 
 + \frac{45}{\sqrt{32\pi^{7}}} \sum_{r_i z>>1} g_i~ (r_i z - z_{\mu_i}) (r_i z)^{3/2} e^{-(r_i z - z_{\mu_i})} \right] \nonumber\\
&  \equiv  \left(\frac{2\pi^2}{45}\right) M^3 z^{-3} g_{*S}(z) ~.
\end{align}

In square bracket one can read the effective entropy degrees of freedom $g_{*S}(z)$. For all practical purposes, the entropy of the non-relativistic sector can be neglected compared to the relativistic one, and away from mass thresholds $g_{*S}(z) \sim \sum_{r_i z<<1} g_i  \sim \text{const}$.

Studying departure from thermal equilibrium it is useful to introduce the \textit{particle number per comoving volume} $Y_i = n_i/s$, which evaluates to

\begin{equation}
Y_i =  \begin{cases} \frac{45 \zeta(3) g_i}{2 \pi^4 g_{*S}(\infty)} \sim \text{const.} & \text{if} \quad T>>m_i,\mu_i ~,\\ \frac{45}{\sqrt{32\pi^7}} \frac{g_i}{g_{*S}(z)} \left( \frac{r_i}{z} \right)^{3/2} e^{-(r_i z - z_{\mu_i})} & \text{if} \quad T<< m_i,\mu_i  ~.\end{cases}
\end{equation}

As a case of interest, let us discuss the relation $\mu= \mu(T)$ that arises when the universe expansion forces a species out of chemical equilibrium. This arises when the temperature drops below the species mass threshold. If only self-interactions are freezing out at the scale of interest, then the entropy of the species is conserved:

\begin{equation}
\frac{d}{dt}(s_i a^3) = 0 = \frac{d}{dt} \left( \frac{m_i-\mu_i}{T} n_i \right) + 3H \left( \frac{m_i -\mu_i}{T} \right) n_i \implies \dot{\mu_i} + (m_i - \mu_i) \frac{\dot{T}}{T} = 0.
\end{equation}

The above equation is solved specifying an initial condition, which can be fixed at the decoupling temperature $T_c$ as $\mu_i(T_c) = \mu_i^{eq}$. Thus we get

\begin{equation}
\mu_i(T) = \mu_i^{eq}\theta( T - T_c) + m_i \left( 1 - \frac{T}{T_c} \right) \theta(T_c - T) ~,
\end{equation}

or, equivalently, in terms of $z$ variables 

\begin{equation}
\mu_i(z) = \mu_i^{eq}\theta( z_c - z) + m_i \left( 1 - \frac{z_c}{z} \right) \theta(z-z_c) ~.
\label{eq:ChemPot}
\end{equation}

Assigning a specific temperature to the decoupling is an approximation since the freezeout happens in a range of temperatures. As we are mostly interested in the asymptotic values of $\mu$ far from $t_c$ this approximation is reasonable.

\section{Relation Between Dark and Visible Sectors}
\label{appendixB}

We assume the SM and dark sector to be efficiently coupled when the latter is deconfined, possibly due to weak interactions. After these interactions freeze out the two sectors decouple and entropies are separately conserved. After this point, their ratio becomes a time-independent constant as well as a measure of relative abundances:

\begin{equation}
\zeta = \frac{s_{SM}}{s_{D}} ~.
\end{equation}

After this point, the temperature of the two sectors evolves differently depending on the degrees of freedom of each sector. We will describe the different phases the system evolves into as the universe expands.

\subsubsection{Deconfined Phase}
As initial conditions, when the DS is deconfined, $T_D \gg \Lambda_D$ as well as  $T_{SM} \gg m_{H}$ the entropies are dominated by relativistic species, so in order to keep $\zeta$ constant, temperatures are linearly related:

\begin{equation}
T_{D} = \left( \frac{g_{*S}^{SM}}{\zeta\,g_{*S}^{DS}} \right)^{1/3} T_{SM} ~.
\end{equation}

Where $g_{*S}^{DS}$ now comprise of dark quarks and gluons. The total energy density driving the universe expansion is

\begin{equation}
\rho_{tot} = \rho_D + \rho_{SM} = \frac{3}{4}[\zeta T_{SM} + T_D ] s_D =   \left[ 1 +\zeta^{4/3} \left( \frac{g_{D}}{g_{SM}} \right)^{1/3} \right] \frac{\pi^2}{30} g_{D}\,T_D^4 ~.
\end{equation}

As a consequence, the universe expands as if only the dark quark-gluon plasma is present, the presence of the SM reflects in a modification of the effective degrees of freedom, and $H \sim \sqrt{\rho} \sim T_D^2$.

\subsubsection{Confinement - during Baryon Freezeout}

When $T_D \sim \Lambda_D$ the dark sector confines breaking chiral symmetry. In this scenario, the dark quark mass is much lower than the confining scale. The dark sector entropy is dominated by relativistic pions. If we assume the SM to be relativistic as well, then the temperature of the two sectors are related by

\begin{equation}
T_{D} = \left( \frac{g_{*S}^{SM}}{\zeta\,g_{*S}^{\pi}} \right)^{1/3} T_{SM} ~.
\end{equation}

This is self-consistent as long as $\Lambda_D \gg m_H$ as expected from a new physics scale. 
The total energy density is found analogously as in the previous case and the scale factor evolves as 

\begin{equation}
H(z_D) = H(\Lambda_D) \left( \frac{T_D}{\Lambda_D} \right)^2 = H(\Lambda_D) \frac{1}{r_{\Lambda}^2 z_D^2} ~,
\end{equation}

where we introduced $r_{\Lambda} = \Lambda_D/M$, with $M$ being the mass scale of the dark matter candidate, in our case the dark baryons. This is the regime in which the dark baryons freezeout and the universe expansion is driven by SM and dark pions. 
For notational convenience we assume $r_{\Lambda} \sim 1$. This is in general not the case, and reintroducing the factors of $r_{\Lambda}$ is straightforward in what follows.
As a concluding remark, number-changing interactions in the pion sector are always active during the baryon freezeout, due to the significant scale hierarchy.

\subsubsection{Freezeout of Number-changing Interactions}

In this regime, the number-changing interactions in the pion sector become relevant for the dark sector thermodynamics. We will restrict our attention to $3 \rightarrow 2$ processes.
Once the expansion of the universe makes the temperature in the dark sector drop below the pion mass $T_D  < m_{\pi}$, the $3\rightarrow 2$ process can freezeout. As long as they are active ($\mu_{\pi_i} = 0$) the entropy of the dark sector will simply be

\begin{equation}
s_D  = \frac{g_{\pi} M^3}{\sqrt{8 \pi^3}}~ r_\pi^{5/2} z_D^{-1/2} e^{-z_D}
\end{equation}

Because of total entropy conservation, integrating from $t_0$ (time at which $T_D = m_{\pi}$) and arbitrary time, we have now

\begin{equation}
\left( \frac{a}{a_0} \right)^3 = (r_\pi z_D)^{1/2}\, e^{z_D-\frac{1}{r_\pi}}  \implies T_D \sim \frac{m_{\pi}}{3 \log \left(a/a_0\right)}  ~,
\end{equation}

In which we assumed $z_D \gg 1$.
The standard model being relativistic, instead has a standard scaling:

\begin{equation}
T_{SM} = T_{0} \left( \frac{a_0}{a} \right) ~.
\end{equation}

Where $T_0$ is the SM at time $t_0$, when $T_D = m_\pi$.
The two temperatures can still be related by the constancy of the entropy ratio:

\begin{equation}
\left(\frac{T_{SM}}{T_D}\right)^3 ~\sim~ z_D^{5/2} e^{-z_D} ~.
\end{equation}

It is then evident that the dark sector is exponentially hotter than the standard model in this phase. To study the universe expansion, it's easier to consider separately the cases in which one sector dominates over the other. The Hubble rate can be computed in this phase to be

\begin{equation}
H(T_D) = \begin{cases}
 H(m_{\pi}) (z_D r_\pi)^{-3/2} e^{-z_D + \frac{1}{r_\pi}} \quad \text{if} \quad \rho_D > \rho_{SM} \\
 H(m_{\pi}) (z_D r_\pi)^{-1/3} e^{- \frac{2}{3}\left(z_D-\frac{1}{r_{\pi}} \right)} \quad \rho_D < \rho_{SM} \\
\end{cases}
\end{equation}

If the $3 \rightarrow 2$ processes are frozen out, one needs to consider the effect of the chemical potential in \eqref{eq:ChemPot}. We then have

\begin{equation}
s_D = \frac{g_\pi M^3 z_c}{\sqrt{8 \pi^3}} e^{-r_\pi z_c} r_\pi^{5/2} z_D^{-3/2}, ~\implies~  T_D = m_\pi \left( \frac{a_0}{a} \right)^2 ~\implies~ T_D \sim T_{SM}^2 ~,
\end{equation}

with $z_c$ corresponds to the $3\rightarrow2$ freeze out temperature, $T_c$. The Hubble rate can be computed in this regime:

\begin{equation}
H(T_D) = \begin{cases}
 H(T_c) \left( \frac{z_c}{z_D} \right)^{3/2} \quad \text{if} \quad \rho_D > \rho_{SM} \\
  H(T_c) \left( \frac{z_c}{z_D} \right) \quad \text{if} \quad \rho_D < \rho_{SM} ~. \\
\end{cases}
\end{equation}

\section{Boltzmann Equation for Number-changing Processes}
\label{appendixC}

In this section we review the derivation of the Boltzmann equation (BE) for number-changing multiparticle processes.
The phase space density for a particle species is $f_i(E)$, it represents the average number of particles in the unit element of phase space. The BE equation then reads schematically as
\begin{equation}
L[f] = C[f], \quad L = E \frac{\partial}{\partial t} - H p^2 \frac{\partial}{\partial E}
\end{equation}
where we used the form of the Liouville operator for the FRW metric, with $H$ being the Hubble rate. The phase space volume element in the box normalization is $d^3p\, gV_c /(2\pi)^3$, where $V_c = a^3 V$. Integrating the Liouville operator on all phase space including the relativistic factor $1/(2E)$ we get
\begin{equation}
\int \frac{d^3 p}{(2\pi)^3} \frac{g}{2 E} \left[ E \frac{\partial f}{\partial t} - H p^2 \frac{\partial f}{\partial E} \right] =  \frac{1}{2} \left[ \dot{n} + 3 H n \right] 
\end{equation}
So that the BE equation takes the form
\begin{equation}
\dot{n} + 3 H n = 2 g \int \frac{d^3 p }{(2\pi)^3} \frac{1}{2E} C[f]\,.
\end{equation}
To understand what the RHS represents, we can see that the LHS is written as
\begin{equation}
\frac{1}{a^3 V}\frac{d}{dt} (n a^3 V) = \frac{1}{V_c} \frac{dN}{dt}\,,
\end{equation}
which is the change in unit time in the number of particles per unit of comoving volume. This is a Lorentz invariant quantity, and the goal is now to write the general collision term for a generic interaction among particle species. \\

All of the states participating the scattering are multiparticle states with occupation number
\begin{equation}
| f_i(E_i), f_j(E_j), ... \rangle\,.
\end{equation}
If the transition operator allows a transition from a particle $i$ to a particle $j$ then it will contain the operator $a^{\dagger}_j a_i$. This will produce a factor $\sqrt{1 \pm f_j} \sqrt{f_i}$ depending on $j$ being a fermion or boson. We will factor these out explicitly, so that our matrix element is between single particle states only. We will also approximate the ensamble to be sufficiently diluted so that $1 \pm f_i \sim 1$ and for equilibrium distributions $f_i^{eq} \sim e^{-\beta(E_i - \mu^{\rm eq}_i)}$.
Consider the following generic structure of the reaction
\begin{equation}
N^{\rm in}_\psi \times \psi + \sum_{i \in I} N_i^{\rm in} \times \phi_i \rightarrow N_\psi^{\rm out} \times \psi + \sum_{i\in O} N_i^{\rm out} \times \phi_i
\end{equation}
The set $I,O$ consist of particle species that can appear at each side of the reaction, with multiplicity $N_i^{\rm in/out}$. As illustrative example consider a Yukawa-$\phi^4$ theory in which we want to study the number changing of fermionic particles $\psi$ due to the reaction
\begin{eqnarray}
\psi + \bar{\psi} + 3 \phi \rightarrow 27 \phi
\end{eqnarray}
We then have $N_\psi^{\rm in} = 1$, $I = \{ \bar{\psi}, \phi\}, N_{\bar{\psi}}^{\rm in} = 1,  N_{\phi}^{\rm in} = 3$ and $O = \{ \phi\}, N_{\phi}^{\rm out} = 27$.
Back to our general reaction, this will correspond to the following collision term for the $\psi$:
\begin{equation}
\dot{n}_{\psi} + 3H n_{\psi} = - \left(N_{\psi}^{\rm in} - N_{\psi}^{\rm out}\right) \sumint \dd \Phi |M_{fi}|^2 \left[ (f_\psi)^{N_\psi^{\rm in}} \prod_{i \in I} (f_i)^{N_i^{\rm in}} - (f_\psi)^{N_\psi^{\rm out}}  \prod_{i \in O} (f_i)^{N_i^{\rm out}} \right]\,,
\end{equation}
where we introduced a short-hand notation for the phase space element
\begin{eqnarray}
\dd \Phi = (2\pi)^4 \delta(p_{\rm tot}^{\rm in} - p_{\rm tot}^{\rm out}) \left[  \prod_{i\in I,\psi} \frac{1}{N_i^{\rm in}!} \frac{1}{2E_i}\frac{\dd^3 p_i}{(2\pi)^3}\right] \left[  \prod_{i\in O,\psi} \frac{1}{N_i^{\rm out}!} \frac{1}{2E_i}\frac{\dd^3 p_i}{(2\pi)^3}\right] \,.
\end{eqnarray}
To avoid further complication, we adopted a slight abuse of notation: same-species particles either in the initial or final state have different momenta. The above definition takes this fact into account implicitly.
Few comments are in order:
\begin{itemize}
\item the factor of $N_{\psi}^I - N_{\psi}^O$ counts how many $\psi$ are lost in a reaction. 
\item We included factors of $1/N_{i}^{\rm out}!$ in the normalization of final states with identical particles. For notational convenience we included them in the definition of the phase space element. This factor is included in the typical cross-section computation in QFT.
\item The factor $1/N_{i}^{\rm in}!$ is necessary as we're integrating over the whole initial phase space, where states with identical particles are present. This is not included in the typical QFT cross section computation.
\end{itemize}
To simplify the expression further we use the fact that during the evolution kinetic equilibrium is always maintained ( the elastic processes are efficient, the expansion is adiabatic) so that 
\begin{equation}
\frac{f_i}{f_i^{eq}} = \, \text{constant independent of $p_i$} \, = \frac{n_i}{n_i^{eq}}\,.
\end{equation}
In what follows we will take $f^{\rm eq}_i = e^{-\beta E_i}$ and incorporate the $\mu_i$-dependent part in the out-of-chemical-equilibrium distributions $n_i$.
The QFT matrix element includes a delta function enforcing energy conservation, which enforces detailed balance:
\begin{equation}
(f^{\rm eq}_\psi)^{N_\psi^{\rm in}} \prod_{i \in I} (f^{\rm eq}_i)^{N_i^{\rm in}} =(f^{\rm eq}_\psi)^{N_\psi^{\rm out}}  \prod_{i \in O} (f^{\rm eq}_i)^{N_i^{\rm out}}.
\end{equation}
Using these facts, the BE equation takes the form
\begin{align}
\dot{n}_{\psi} + 3H n_{\psi} =& - \left(N_{\psi}^{\rm in} - N_{\psi}^{\rm out}\right) \sumint \dd \Phi |M_{fi}|^2 (f^{\rm eq}_\psi)^{N_\psi^{\rm in}} \prod_{i \in I} (f^{\rm eq}_i)^{N_i^{\rm in}} \\ \nonumber
& \times \left[ \left(\frac{n_\psi}{n_\psi^{\rm eq}}\right)^{N_\psi^{\rm in}} \prod_{i \in I} \left(\frac{n_i}{n_i^{\rm eq}}\right)^{N_i^{\rm in}} - \left(\frac{n_\psi}{n_\psi^{\rm eq}}\right)^{N_\psi^{\rm out}}  \prod_{i \in O} \left(\frac{n_i}{n_i^{\rm eq}}\right)^{N_i^{\rm out}} \right]\,.
\end{align}
Switching to the commonly used variables $Y_i = n_i/s$ this is recasted in the more familiar form
\begin{align}
\label{eq:generalBE}
\dot{Y}_{\psi} =& - \left(N_{\psi}^{\rm in} - N_{\psi}^{\rm out}\right) \frac{\Gamma_{ij}(z)}{J(z)} \, s^{N_\psi^{\rm in} + \sum_{i \in I} N_i^{\rm in}-1} \\ \nonumber
&\times \left\{ Y_\psi^{N_\psi^{\rm in}}\prod_{i\in I} Y_i^{N_i^{\rm in}} - \left[ \left( \frac{Y_\psi}{Y_\psi^{eq}} \right)^{N_\psi^{\rm out}} \prod_{i \in O} \left( \frac{Y_i}{Y_i^{eq}} \right)^{N_i^{\rm out}}\right] \left[ (Y_\psi^{\rm eq})^{N_i^{\rm in}}\prod_I (Y_i^{eq})^{N_i^{\rm in}} \right]\right\}\,.
\end{align}
The decay factor according to 
\begin{equation}
\label{eq:generalDecayW}
\Gamma_{ij}(z) = \frac{1}{(n_\psi^{\rm eq})^{N_\psi^{\rm in}}\prod_{i \in I} (n_i^{\rm eq})^{N_i^{\rm in}} }  \sumint \dd \Phi|M_{fi}|^2 (f^{\rm eq}_\psi)^{N_\psi^{\rm in}} \prod_{i \in I} (f^{\rm eq}_i)^{N_i^{\rm in}}\,.
\end{equation}
The function $J(z) = \dd z / \dd t $ is the jacobian factor coming form the variable change $t\rightarrow z$.

\subsubsection{Calculation of Jacobian Factors}

In this section we compute the Jacobian factor that appears in the Boltzmann equation expressed in term of dark temperature:

\begin{equation}
J(z_D) =  \frac{dz_D}{dt} = - \frac{1}{T_D} \frac{dT_D}{dt} z_D ~.
\end{equation}

In the regime in which the two-sector temperature are linearly related and the dark sector is dominated by relativistic pions, we have

\begin{equation}
0 =  \frac{d}{dt} \left(  T_D^3 a^3 \right) \implies \frac{1}{T_D}\frac{dT_D}{dt} = - H(T_D) ~,
\end{equation}

so that $J(z_D) = H(z_D) z_D$. When the pions are non-relativistic and number-changing processes are active, we have instead

\begin{equation}
0 = \frac{d}{dt} ( a^3 z_D^{-1/2} e^{-z_D} ) \implies J(z_D) = \frac{d z_D}{dt} = \frac{6 z_D H(z_D)}{1 + 2z_D} ~.
\end{equation}

Finally, once these processes are decoupled we get

\begin{equation}
0 = \frac{d}{dt} ( a^3 z_D^{-3/2} ) \implies J(z_D) = \frac{d z_D}{dt} = 2 z_D H(z_D) ~.
\end{equation}

\phantom{.}



\footnotesize
\bibliographystyle{abbrv}



\end{document}